\RequirePackage{setspace}
\PassOptionsToPackage{bookmarks=false}{hyperref}
\documentclass[fleqn,usenatbib]{mnras}
\pdfoutput=1
\usepackage{amsmath}
\usepackage{multirow}
\usepackage{graphicx}
\usepackage[authoryear]{natbib}
\usepackage{amssymb,txfonts}
\usepackage{lscape} 
\usepackage{adjustbox}
\usepackage{setspace}
\usepackage{hyperref}

\newcommand{\nicer}{\textit{NICER }}
\newcommand{\nustar}{\textit{NuSTAR}}

\newcommand{\swift}{{\it Swift }}

\newcommand{\maxi}{MAXI~J1348-630}

\title[\textit{NuSTAR} spectral study of \maxi]{\textit{NuSTAR} monitoring of \maxi: evidence of high density disc reflection}

\author[S. Chakraborty et al.]{
Sudip Chakraborty$^{1}$\thanks{E-mail: sudipchakraborty93@gmail.com},
Ajay Ratheesh$^{2,3}$,
Sudip Bhattacharyya$^{1}$,
John A. Tomsick$^{4}$,
\newauthor
Francesco Tombesi$^{2,5,6,7}$, 
Keigo Fukumura$^{8}$,
Gaurava  K.  Jaisawal$^{9}$\\
\\
$^{1}$Department of Astronomy and Astrophysics, Tata Institute of Fundamental Research, 1 Homi Bhabha Road, Mumbai 400005, India\\
$^{2}$Department of Physics,  Tor Vergata University of Rome,
via della Ricerca Scientifica 1, 00133, Rome, Italy.\\
$^{3}$INAF-IAPS, Via del Fosso del Cavaliere 100, I-00133 Rome, Italy\\
$^{4}$Space Sciences Laboratory, University of California Berkeley, 7 Gauss Way, Berkeley, CA 94720-7450, USA\\
$^{5}$INAF - Astronomical Observatory of Rome, Via Frascati 33, I-00078 Monte Porzio Catone (Rome), Italy\\
$^{6}$Department of Astronomy, University of Maryland, College Park, MD 20742, USA\\
$^{7}$NASA/Goddard Space Flight Center, Greenbelt, MD 20771, USA\\
$^{8}$James Madison University, 800 South Main Street, Harrisonburg, Virginia 22807, USA\\
$^{9}$National Space Institute, Technical University of Denmark, Elektrovej 327-328, DK-2800 Lyngby, Denmark
}

\date{ Accepted 2021 September 4. Received 2021 September 4; in original form 2021 April 25}

\pubyear{2021}

\begin{document}
\label{firstpage}
\pagerange{\pageref{firstpage}--\pageref{lastpage}}
\maketitle

\begin{abstract}

We present the broadband spectral analysis of all the six hard, intermediate and soft state \textit{NuSTAR} observations of the recently discovered transient black hole X-ray binary MAXI~J1348-630 during its first outburst in 2019. We first model the data with a combination of a  multi-colour disc and a relativistic blurred reflection, and, whenever needed, a distant reflection. We find that this simple model scheme is inadequate in explaining the spectra, resulting in a very high iron abundance. We, therefore, explore the possibility of reflection from a high-density disc. We use two different sets of models to describe the high-density disc reflection: \textsc{relxill}-based reflection models, and \textsc{reflionx}-based ones. The \textsc{reflionx}-based high-density disc reflection models bring down the iron abundance to around the solar value, while the density is found to be $10^{20.3-21.4} \rm cm^{-3}$. We also find evidence of a high-velocity outflow in the form of $\sim$7.3 keV absorption lines. The consistency between the best-fit parameters for different epochs and the statistical significance of the corresponding model indicates the existence of high-density disc reflection in MAXI~J1348-630.

\end{abstract}


\section{Introduction}
Black hole X-ray binaries (BHBs) are systems with a black hole and a companion star. BHBs can be classified into persistent and transients based on the magnitude of flux change seen in them. In transient BHBs, the flux in the X-ray band can change by few orders of magnitude. The primary spectrum of BHBs can be modelled with a multi-temperature black body radiation emitted by an accretion disc \citep{Novikov_1973blho.conf..343N,Shakura_1973A&A....24..337S}, a Comptonisation (power-law) component due to the up-scattering of the soft disc photons in a high temperature compact structure called corona \citep{Sunyaev_1_1980A&A....86..121S,Sunyaev_2_1985A&A...143..374S}, and a reflection of the Comptonized photons off the top layer of the disc. The reflection features consist of iron emission lines at 6 to 7 keV along with a Compton hump at approximately 30-50 keV and an excess below 1 keV\citep{Fabian_2016}, further smeared by the relativistic effects in the vicinity of the black hole. The relativistic  reflection  component  provides insights on the  disc-corona  geometry (such  as  the  size of the coronal region  and the inner radius of the disc) and the changes of inner accretion processes throughout the evolution of BHB spectral states. Thus the BHB spectral states are characterised by the geometry of the disc and corona, and the interplay between the direct and reflection components. The main spectral states of transient BHBs are the low hard state (LHS) and the high soft state (HSS) \citep{Belloni_2000A&A...355..271B}. The main conjecture is that in the HSS the disc extends until inner most circular orbit (ISCO), while the disc is truncated at a larger distance in the LHS \citep{Esin_1997,Belloni_2000A&A...355..271B,Ponti_2012}. While transiting between these states, they undergo a q-shaped hysteresis within the hardness intensity diagram (HID) \citep{Fender_2004,Remillard_2006ARA&A..44...49R}. The source rises in intensity from the LHS before transiting into the HSS, and then falls in intensity before transiting back into the low hard state before vanishing into quiescence. Outflows in the form of jet and winds are also observed in these systems. A jet is generally seen in the hard state and stronger while the source transits from soft to hard state \citep{Fender_2004}. Even though disc winds are stronger in the soft state in comparison to the hard state \citep{Neilsen_2012MNRAS.422L..11P,Fukumura_2021,Ratheesh_2021}, the state dependence of disc wind is not clearly understood.

A common outcome of reflection modelling of BHB X-ray spectra  is  super-solar iron  abundance \citep{Parker_2015,Walton_2016}. This high iron abundance is a feature seen in AGN as well \citep[e.g., ][]{Parker_2018}. Radiative levitation, or the enhancement of the metallicity due to radiation-pressure dominance of the inner disc, has been proposed as an explanation behind these super-solar abundances \citep{Reynolds_2012}. While in some AGN, this kind of high metallicity could indeed be real \citep{Wang_2012}, in most of the cases, this is perhaps an artefact of the model and the assumptions therein.
An alternate  explanation  of  the  high  iron  abundances  involves reflection off a high-density accretion disc. For most of the widely used disc reflection models, a  constant  electron  density of ($n_{\rm e}=10^{15}\rm{cm^{-3}}$, inspired by the values observed in AGN) is assumed for the  top  layer  of  the  disc, along with the assumption that the actual density does not have a significant effect on the predicted spectrum.
While this density  is  appropriate  for the very highest mass supermassive BHs in AGN  \citep[e.g.,][]{Grupe_2010, Jiang_2019b}, BHBs may require higher densities \citep{Garcia_2016}.
However, the free–free absorption and heating of the disc have both found to be quadratically dependent on density. As the density increases, the rise in free–free absorption leads to an increase in the temperature of the top layer of the disc, causing extra thermal emission and increasing the reflected  emissions  below  1  keV \citep{Ross_2007,Garcia_2016}.  Thus, models taking into account this effect has been demonstrated to relieve  the  very  high  iron abundance required in other reflection models \citep[e.g.,][]{Tomsick_2018,Jiang_2019b}.\\

MAXI J1348-630 is such a transient BHB discovered by Gas Slit Camera (GSC) onboard Monitor of All-sky X-ray Image ($MAXI$) \citep{Maxi_2009PASJ...61..999M}, and classified as a BHB based on the estimated mass and spectral features \citep{Tominaga_2020}. Previous studies, using $MAXI$/GSC, $Swift$/XRT \citep{Tominaga_2020} and $NICER$ data \citep{zhang_2020MNRAS.499..851Z}, indicate that MAXI J1348-630 also exhibits a q-shaped hysteresis and that it is most likely a BHB. The source exhibited a hard to soft transition from MJD 58517 to MJD 58530 and returned back to the hard state around MJD 58600 \citep{Tominaga_2020}. The distance to the source was estimated to be 3.39 kpc, based on the precise measurements using a dust scattering ring around the source (utilizing $SRG$/eROSITA and $XMM$-Newton data; \citet{Lamer_dustRing_2020}). Monitoring of the inner disc radius from the spectral analysis indicate a black hole mass of 11 $\pm$ 2 M$_{\odot}$ \citep{Tominaga_2020,Lamer_dustRing_2020}. Even during the soft state the inner disc temperature was in the range of 0.5 to 0.7 keV \citep{Tominaga_2020,zhang_2020MNRAS.499..851Z}. The source underwent a failed outburst without hard to soft transition as also seen previously in some transients BHBs \citep{Tominaga_2020,Homan_2013ApJ...775....9H,furst_2015ApJ...808..122F,Stiele_2020ApJ...889..142S}. The source showed type B quasi periodic oscillations (QPOs) during hard to soft transition generally accompanied by a relativistic jet \citep{Belloni_2020}. The source was also detected in optical and ultraviolet (UV) and radio frequencies \citep{Denisenko_2019ATel12430....1D,Kennea_2019ATel12434....1K,Russel_2019ATel12829....1R,Russel_radio_2019ATel12456....1R}.   \\

In this present investigation, we provide a broadband spectral view of MAXI~J1348-630, observed with the Nuclear Spectroscopic Telescope Array (\textit{NuSTAR}, \citep{Harrison_2013}) during its first complete outburst during January-June, 2019. 
With its pile-up free performance, reasonably good energy resolution ($\sim$400 eV at 10 keV) and sensitivity, \textit{NuSTAR} provides an opportunity to study the relativistic reflection features along with a well-constrained inner disc radius and inclination with great precision. We use all the 6 \textit{NuSTAR} data (see table ~\ref{tab:obs}) during the first outburst of \maxi, spanning the hard, intermediate and soft states, and model the spectra with different disc reflection models. In particular, we focus on the possibility of high density reflection. The paper is organised as follows. In section \ref{sec:obs}, we describe the data reduction procedure of \textit{NuSTAR}, as well as the $MAXI$/GSC and \swift/BAT lightcurves and HID of \maxi. We then present an in-depth spectral analysis of \textit{NuSTAR} data in section ~\ref{sec:w_refl}, first with regular density disc reflection models and then with high density reflection models in section ~\ref{sec:relxillD} and ~\ref{sec:reflionxhd}. Finally, we summarise our results and discuss their implications in section \ref{sec:discussion}.


\section{Observations and Data reduction}

\subsection{$MAXI$/GSC and \swift/BAT Lightcurve and HID} \label{sec:lc}
\begin{figure}
\begin{centering}
\includegraphics{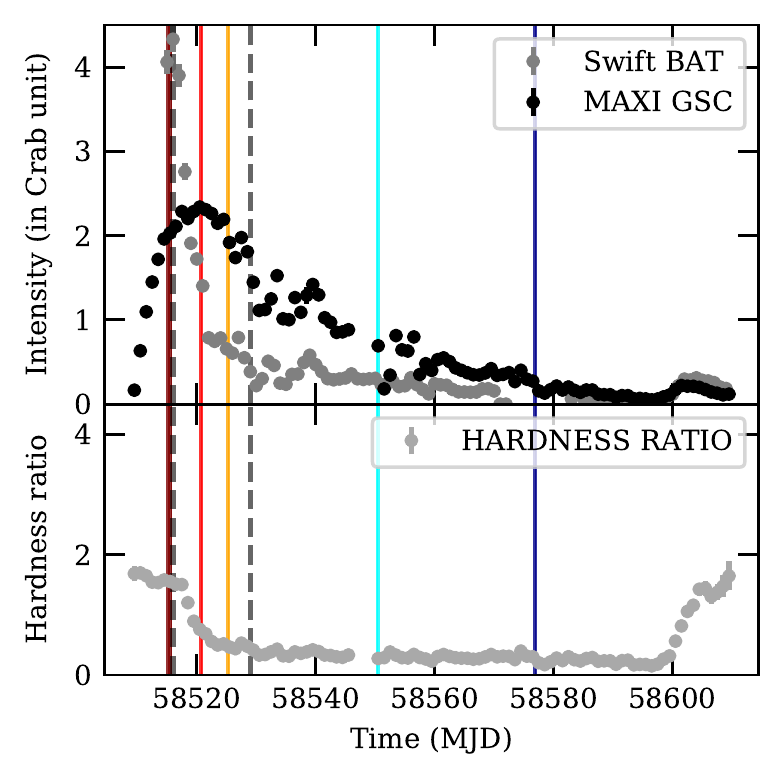}
\par\end{centering}
\caption{Top: $MAXI$/GSC (4-10 keV), $Swift$/BAT (15-50 keV) lightcurves are plotted in black and grey. Bottom: The corresponding hardness ratio (10-20 keV/4-10 keV) from $MAXI$/GSC. All the observations used in this analysis are marked by vertical colour shaded regions. The dotted lines indicate the start and stop of the hard to soft state transition.
}
\label{maxi_bat_lc}
\end{figure}

Since the BHBs evolve in timescales of the order of seconds to days, it is important to identify the state of the source based on continuous monitoring of the source. Daily monitoring of the X-ray sky have been possible due to the Neil Gehrels $Swift$ Observatory \citep[\swift/BAT:][]{SwiftBAT_2013ApJS..209...14K} in the hard X-ray band and Gas Slit Camera on board Monitor of All-sky X-ray Image in the soft X-ray band \citep[MAXI/GSC:][]{Maxi_2009PASJ...61..999M}. The daily averaged lightcurves of $Swift$/BAT and $MAXI$/GSC have been obtained from \url{https://swift.gsfc.nasa.gov/results/transients/} and \url{http://maxi.riken.jp}. The top panel of Fig. \ref{maxi_bat_lc} shows the $Swift$/BAT (grey) and $MAXI$/GSC (black) crab normalized lightcurves in 15-50 keV and 4-10 keV energy range, and the bottom panel shows the hardness ratio in 4-10 keV and 2-4 keV ranges. 1 Crab unit corresponds to 1.1 $\times$ 10$^{-8}$ ergs s$^{-1}$ cm$^{-2}$ and 9.2 $\times$ 10$^{-9}$ ergs s$^{-1}$ cm$^{-2}$ in the 4-10 keV and 2-4 keV energy bands covered by $MAXI$/GSC, and 1.3 $\times$ 10$^{-8}$ ergs s$^{-1}$ cm$^{-2}$ in the 15-50 keV energy band covered by \textit{\swift/BAT} \citep{Kirsch_2005}. The dotted lines indicate the start and stop of the hard to soft state transition obtained from \cite{Tominaga_2020}. The vertical coloured lines indicate the \textit{NuSTAR} observations used in this analysis as outlined in the subsection below (also refer to table \ref{tab:obs}). Fig. \ref{maxi_hid}, shows the hardness intensity diagram (HID). The soft and hard bands used for the hardness ratio is as same as the bottom panel of Fig. \ref{maxi_bat_lc}. The HID indicates that the source underwent a q-shaped hysteresis starting from the low hard state as generally seen in black hole X-ray binaries. The coloured points in the HID indicate the $MAXI$/GSC data on the days of the observations used in this work.

\begin{table*}
\centering
\begin{tabular}{ |c|c|c|c|c|c| } 
\hline
Instrument & Obs ID & Obs. date  & Exposure (s) & Abbreviation & Epoch\\
        & & (yyyy-mm-dd)   &  & &\\
\hline
\nustar & 80402315002 & 2019-02-01 & 3038 & Nu02 (maroon) & E1 \\
\nustar & 80402315004 & 2019-02-01 & 736 & Nu04 (maroon) & E1 \\
\nustar & 80402315006 & 2019-02-06 & 4520 & Nu06 (red) & E2 \\
\nustar & 80402315008 & 2019-02-11 & 4639 & Nu08 (orange) & E3 \\
\nustar & 80402315010 & 2019-03-08 & 9714 & Nu10 (cyan) & E4 \\
\nustar & 80402315012 & 2019-04-03 & 12490 & Nu12 (darkblue) & E5 \\
\hline
\end{tabular}
\caption{\textit{NuSTAR} observation details for \maxi. We include only the observations considered in this work. The colours, mentioned within the brackets, indicate the corresponding observations in all the figures. The colours are in accordance with matplotlib colour palette.}
\label{tab:obs}
\end{table*}

\subsection{\nustar} \label{sec:obs}
\label{sec:nustar} 
\begin{figure}
\begin{centering}
\includegraphics[width=\linewidth]{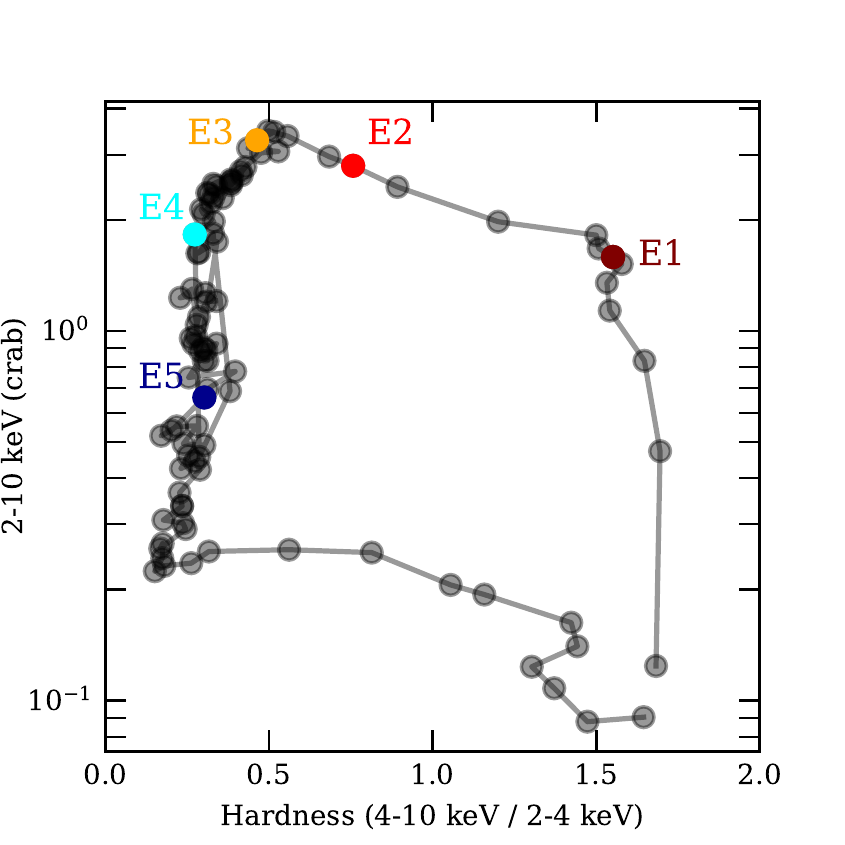}
\par\end{centering}
\caption{The hardness intensity diagram (HID) for MAXI~J1348-630. For the hardness ratio 4-10 keV and 2-4 keV energy band in $MAXI$/GSC is taken as hard and soft bands respectively (in grey). The coloured points indicate the $MAXI$/GSC data during the $NuSTAR$ observations used in this work. }
\label{maxi_hid}
\end{figure}

MAXI~J1348-630 was observed with \textit{NuSTAR} \citep{Harrison_NuSTAR_2013} 6 times in 5 different epochs during its first complete outburst cycle in January-June, 2019. The details of the \textit{NuSTAR} observations and their colour coding followed throughout the rest of this work, are presented in table~\ref{tab:obs}. From figure~\ref{maxi_hid}, it can be seen that the epoch E1 (maroon data point; comprised of the two contemporaneous \textit{NuSTAR} observations Nu02 and Nu04, as denoted in table~\ref{tab:obs}) is located in the hard state, while the epoch E2 (red; comprised of the \textit{NuSTAR} observation Nu06) is situated at the hard-soft transition in the HID. The epochs E3-E5 (orange, cyan and darkblue points in figure~\ref{maxi_hid}) all belong to the soft state, with decreasing flux levels. The \textit{NuSTAR} data are processed using  v2.0.0 of the NuSTARDAS pipeline. We also use \textit{NuSTAR} CALDB v20200813. 
A rip in the Multi Layer Insulation (MLI) at the exit aperture of the $NuSTAR$ Optics Module A (OMA), aligned with detector focal plane module FPMA, occurred presumably in early 2017. This has resulted in increased photon fluxes through OMA, resulting in low energy ($<$10 keV, more prominent in softer energies) excesses in FPMA spectra, as compared to FMPB. This effect of reduction in MLI covering fraction has been taken care of by implementing (through the \textit{numkarf} module in NuSTARDAS v2.0.0, as well as the latest CALDB) time/temperature dependent correction to the FPMA effective area using specific on-axis ARF files \citep{Madsen_2020}. However, there exist pathological observations in which this correction falls short. For these individual cases, the $NuSTAR$ SOC has provided an XSPEC multiplicative table model\footnote{\url{http://nustarsoc.caltech.edu/NuSTAR_Public/NuSTAROperationSite/mli.php}}. We come across one such observation (Nu10/E4), and implement the suggested correction (see section~\ref{sec:w_refl}).
We filter background flares due to enhanced solar activity by setting saacalc = 2, saamode = OPTIMIZED, and tentacle = no in NUPIPELINE. Furthermore, due to the source being extremely bright, we used a modified value of the 'statusexpr' keyword in NUPIPELINE, statusexpr="STATUS==b0000xxx00xxxx000", as suggested in the HEASARC \textit{NuSTAR} analysis guide\footnote{\url{https://heasarc.gsfc.nasa.gov/docs/nustar/analysis/}}. The source spectra are extracted from circular regions of the radius 120$^{''}$ centred on the source location. To avoid contamination by source photons from the extremely bright source regions, the background spectra are extracted from blank regions on the detector furthest from the source location.
The spectra are grouped in $ISIS$ \citep{Houck_2000} version 1.6.2-41 to have a signal-to-noise ratio of at least 25 per bin (15 in case of the fainter spectra in E5), to facilitate $\chi^2$ fitting statistics.

\section{Data Analysis and Results} \label{sec:results}
The spectral fitting and statistical analysis are carried out using the XSPEC version v-12.11.0 \citep{Arnaud_1996}. For the joint fitting between FPMA and FPMB, a cross-normalisation constant (implemented using \textsc{Constant} model in XSPEC) is allowed to vary freely for FPMB and is assumed to be unity for FPMA. To avoid the unexplained residuals below 4 keV in some of the observations, possibly related to the issue with the MLI blanketing as detailed in \citet{Madsen_2020}, an energy range between 4 keV and 79 keV is considered for the spectral fittings of all the epochs. To avoid the sharp instrumental features (as reported by \cite{Xu_J1535_2018}), energies between 11-12 keV and 26-28 keV are excluded. All the models, as described below, include the Galactic absorption through the implementation of the \textsc{TBabs} model. The corresponding abundances are set in accordance with the \cite{Wilms_2000} photoelectric cross-sections. The Galactic neutral hydrogen column density ($N_{\rm H}$) is fixed to $8.6 \times 10^{21} \ \rm cm^{-2}$ \citep{Tominaga_2020}  for all the described models. All parameter uncertainties are reported at the 90\% confidence level for one parameter of interest. In addition to the $\chi^2$ and the degrees of freedom, we also provide the null hypothesis probability for each of the models we use. In order to facilitate comparisons between the different nested and non-nested models and to select better models, we also state the Akaike Information Criterion \citep[AIC, ][]{Akaike_1974} and Bayesian Information Criterion \citep[BIC, ][]{Schwarz_1978}. AIC is more suitable to compare non-nested models, whereas BIC is more suited when nested models are compared \citep{Kass_1995}, although that is not a necessity. In general, BIC penalises models with higher number of free parameters more severely. The AIC is defined as:
\begin{equation}
\label{aic}
\rm AIC = -2\ln\mathcal{L}+2k
\end{equation}
where $\mathcal{L}$ is the likelihood of the best-fit model, $k$ is the number of free parameters. The BIC, on the other hand, is defined as:
\begin{equation}
\label{bic}
\rm BIC = -2\ln\mathcal{L}+k\ln(N)
\end{equation}
where $N$ is the total number of data points. The values of the $\chi^2$ for the best-fit models, $\chi^{2}_{\rm min}$, in our fits can be converted to likelihood by equating the two: $\chi^{2}_{\rm min}=-2\ln\mathcal{L}$. The values of AIC and BIC thus calculated for all our best-fit models, are provided in Tables ~\ref{tab:M1},~\ref{tab:M2},~\ref{tab:M3} and ~\ref{tab:M4}. The model with the lowest AIC or BIC is the most preferred one.
It should be noted, however, that the AIC and BIC compares models from pure statistical perspective; and one should also consider the physical interpretation of models and the reasonable range of their parameter values while selecting the `better' model.

\begin{figure*}
\centering
	\includegraphics{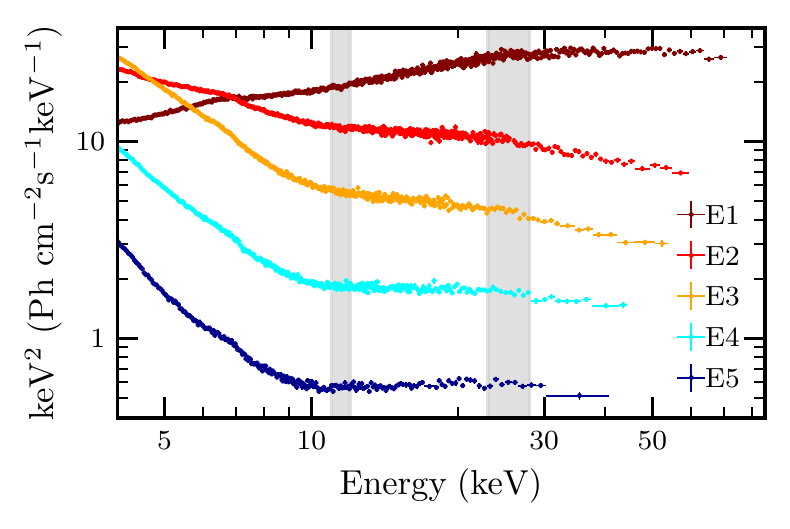}\quad
	\includegraphics{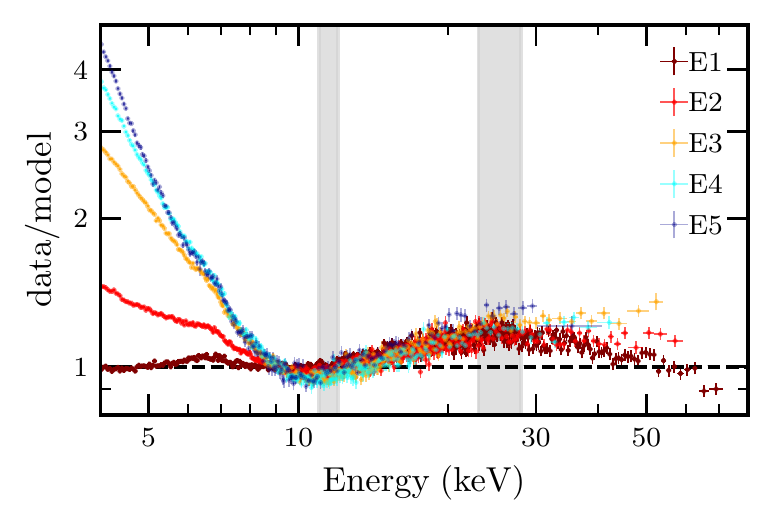}
  	\caption{Left panel: \textit{NuSTAR} spectra, unfolded with a \textsc{constant} model to see the spectral variations between the epochs. The colour coding for each epoch is detailed in table ~\ref{tab:obs}. 
  	 Right panel: \textit{NuSTAR} residuals, in form of data/model ratio, after a power-law fit. The soft excess, broad iron line and the Compton hump can be noticed. See section ~\ref{sec:results} for further details. The shaded grey region consists of energy bands ignored for detailed spectral fitting. }
    \label{fig:eeufspec}
\end{figure*}

The spectra from all the epochs (E1 to E5) considered here are shown unfolded to a constant model in Figure ~\ref{fig:eeufspec} (left panel). A consistent spectral softening from hard to soft state can be observed. With our signal-to-noise ratio based binning, we find that the epochs E4 and E5 possess no statistically significant energy bins above 50 keV.

To highlight the spectral features more clearly, we plot the spectra from all the epochs as ratios to the best-fitting power-law model in Figure ~\ref{fig:eeufspec} (right panel). For the powerlaw fits, we only consider the energy intervals of 8-11 and 40-79 keV, where reflection from the disc has minimal effect. We then fit the 8-11,40-79 keV spectrum of each epoch with an absorbed cutoff power-law model, \textsc{TBabs$\times$cutoffpl} in XSPEC notation. Finally, we generate the 4-79 keV residuals by diving the data from each epoch with the corresponding absorbed cutoff power-law model. The resulting figure ~\ref{fig:eeufspec} shows a broad Fe K-$\alpha$  emission line peaking around 6.5 keV and a Compton hump at $\sim$20-40 keV for all the epochs, hinting the presence of relativistic reflection expected from an accretion disc extending close to a black hole \citep{Fabian_2000}.

\subsection{Quantitative modelling with reflection} \label{sec:w_refl}

For a detailed investigation of the broadband spectra including the reflection features, we start with the self-consistent relativistic disc reflection models from \textsc{relxill} model suite (relxill v1.3.10 : \cite{Dauser_2014}, \cite{Garcia_2014}). We assume an extended corona and use the model \textsc{relxillCp} which internally includes an thermal Comptonisation (\textsc{Nthcomp}: \cite{Zdziarski_1996},\cite{Zycki_1999}) continuum. Due to the strong degeneracy between the inner radius ($R_{\rm in}$) of the accretion disc, and the dimensionless black hole spin parameter ($a$), we fit the data for $R_{\rm in}$ assuming a maximally spinning black hole ($a = 0.998$). We fix the outer edge of the accretion disc ($R_{\rm out}$) at the maximum value for the model of $400R_{\rm g}$ (where $R_{\rm g}$ is the gravitational radius of the black hole, defined as $R_{\rm g}=GM/c^2$) and the emissivity indices at 3 ($q_1=q_2=3$). On the other hand, we keep the inclination angle free. To account for the narrow core of the Fe-K$\alpha$ line in the hard state data (Epoch E1), we use the unblurred reflection model \textsc{xillverCp} \citep{Garcia_2010}. We use \textsc{xillverCp} only as a reflection component, freezing the refl\_frac of \textsc{xillverCp} at -1, as only insignificant variations are found in the subsequent fits if the refl\_frac is allowed to vary freely. We require an additional \textsc{xillverCp} for E3 as well. For both the cases, we assume the \textsc{xillverCp} component to be neutral to maintain simplicity. The electron temperatures ($T_{\rm e}$), spectral indices ($\Gamma$), iron abundances ($A_{\rm Fe}$) and inclinations ($\theta$), are assumed to be equal between the \textsc{relxillCp} and \textsc{xillverCp} components.
To model the soft X-ray excess (as evident from Figure ~\ref{fig:eeufspec}), we use the  multicoloured disc black-body (\textsc{diskbb}: \cite{Mitsuda_1984}, \cite{Makishima_1986}) model. 
As suggested by the instrument team, we also account for the MLI correction for FPMA in the pathological case of Nu10 (Epoch E4). This is done by the introduction of the multiplicative model \textsc{nuMLIv1.mod}\footnote{\url{http://nustarsoc.caltech.edu/NuSTAR_Public/NuSTAROperationSite/mli.php}} suggested by \textit{NuSTAR} SOC \citep{Madsen_2020}, and fixing the MLI covering fraction of FPMA to the recommended value of 0.83. 

After the continua of all the epochs are modelled, absorption features around 7.2-7.3 keV are observed in the residuals for all of the 5 epochs. This kind of absorption features in the Fe-K band are commonly associated with blueshifted FeXXV/FeXXVI lines, and are believed to  arise  from  absorption  by  outflowing  material  launched from the accretion disc \citep[e.g., ][]{Ponti_2012}. The  narrow  absorption  line complex cannot be properly resolved by $NuSTAR$, hence we model it simply using a Gaussian absorption line model (\textsc{gabs} in XSPEC). 

This constitutes our model M1 in Table ~\ref{tab:M1}:\\

Nu02/04/08:\\
\verb'TBabs'$\times$\verb'gabs'$\times$\verb'(diskbb+relxillCp+xillverCp)'\\

Nu06/10/12:\\
\verb'TBabs'$\times$\verb'gabs'$\times$\verb'(diskbb+relxillCp)'\\

For epoch E1, the two \textit{NuSTAR} observations (Nu02 and N04) are carried out within a short interval. For a consistent fit between the two observations, we allow the continuum parameters (Innermost temperature ($T_{\rm in}$) of the \textsc{diskbb} component, $T_{\rm e}$ and $\Gamma$ of the \textsc{Nthcomp} component in \textsc{relxillCp}) and all the normalisations to vary freely between them, tying up the absorption and reflection parameters ($R_{\rm in}$, $A_{\rm Fe}$, $\theta$ and ionisation parameter\footnote{Ionisation parameter is defined as $\xi=4\pi \rm F_X/n$, where $F_X$ is the ionising continuum flux and $n$ is the gas density} ($\xi$) of the \textsc{relxillCp} component) between the two observations. 

From Table ~\ref{tab:M1}, we find that as the source moves from hard to soft states (E1 to E5), the $T_{\rm in}$ first increases before decreasing again, while the $T_{\rm e}$ increases before getting pegged at the parameter upper bound for \textsc{relxillCp} model. The disc is found to be highly ionised, and the inner edge of accretion disc is found to contract from $\sim 10 R_{\rm g}$ to near the innermost stable circular orbit (ISCO). The inclination is found to be around 30-45 degrees. 

\begin{figure}
\centering
    \includegraphics{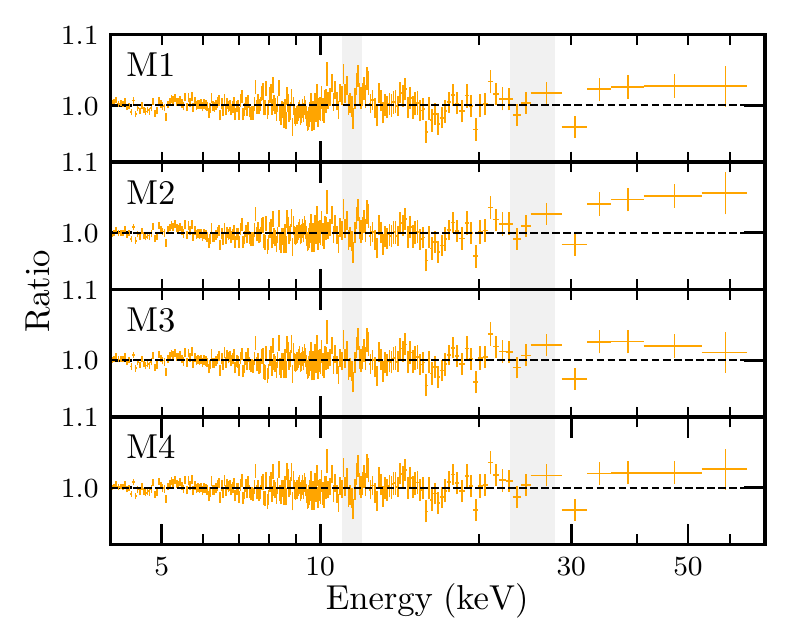}
    \includegraphics{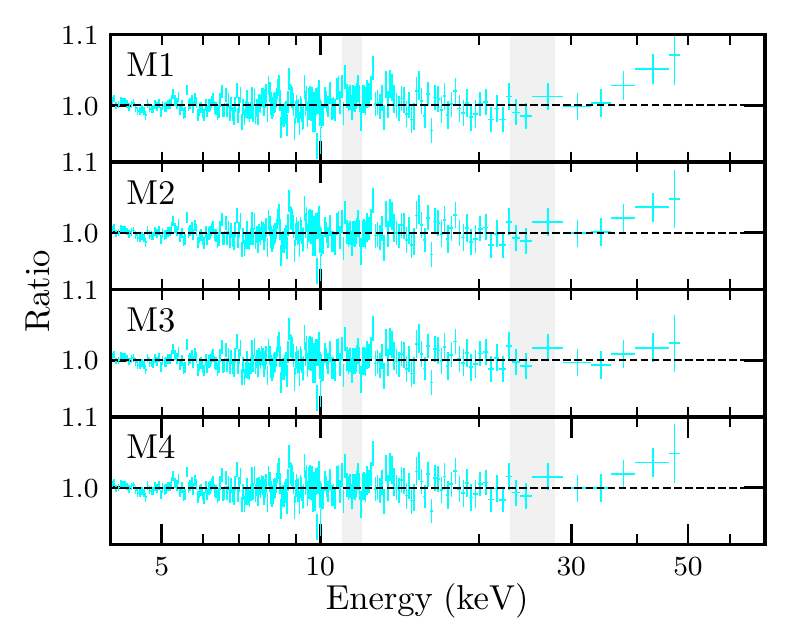}
  	\caption{Residuals, in the form of data/model ratios, for each of the 4 models for two representative epochs E3 (orange, top panel) and E4 (cyan, bottom panel). 	For better visual representation, we display only the FPMA residuals. The light grey shaded regions indicate the 11.0-12.0 and 26.0-28.0 keV regions, ignored in this work due to unexplained systematics. See section~\ref{sec:w_refl} for model descriptions.}
    \label{fig:ratio}
\end{figure}

However, there are some problems with the best-fit M1 model, as apparent from Table ~\ref{tab:M1}. First of all, the iron abundance of the disc is found to be significantly higher than solar, sometimes pegging at the upper limit of 10 times solar. This fact is further verified by performing a Markov chain Monte Carlo (MCMC) analysis with 50 walkers with chain lengths of 5000 for E2, E4 and E5. 
These values are most likely overestimated, given the ubiquity of apparent super-solar iron abundances \citep{Garcia_2018}. The overestimate has previously been attributed to high density discs \citep{Svensson_1994,Garcia_2016} which would show stronger iron lines at a given metallicity \citep{Garcia_2016,Tomsick_2018}. Secondly, this unusually high iron abundance is also found to be sometimes accompanied by systematically increasing residuals above 50 keV (see the uppermost panels showing the residuals in E3 and E4, in Figure ~\ref{fig:ratio}). This feature has also been observed for Cyg X-1 by \citet{Tomsick_2018}. 
To address these issues, we subsequently explore the possibility of a high density disc by modeling all the \textit{NuSTAR} spectra with high density reflection models.

\subsubsection{relxill-based high density reflection models} \label{sec:relxillD}

To extend the disc density from its constant value of $n_{\rm e}=10^{15}cm^{-3}$ to higher values, we first use the \textsc{relxill}-based high density reflection model, \textsc{relxillD}. In this model, the continuum is assumed to be a power-law with a high energy cutoff (\textsc{cutoffpl} in XSPEC), with the cutoff energy ($E_{\rm cut}$) fixed at 300 keV. Instead, the disc density is treated as a free parameter, and can range from $10^{15}cm^{-3}$ to a maximum value of $10^{19}cm^{-3}$. In case of E1 and E3, similar to M1, we add a neutral distant reflection as the best-fit model. To make the model consistent, we use the cutoff power-law version of the distant reflection component, \textsc{xillver}. All the parameters are treated in a similar manner as M1. Thus our model M2 (see Table ~\ref{tab:M2}) becomes:
Nu02/04/08:\\
\verb'TBabs'$\times$\verb'gabs'$\times$\verb'(diskbb+relxillD+xillver)'\\

Nu06/10/12:\\
\verb'TBabs'$\times$\verb'gabs'$\times$\verb'(diskbb+relxillD)'\\

From Table ~\ref{tab:M2} it can be seen that for the same degrees of freedom, best-fit M2 provides $\Delta \chi^2$ of 18-23 less than M1 for most of the epochs. For the hard state observations in E1, the fit gets worse due to $kT_{\rm e}$ being fixed at a high value of 300 keV. The iron abundances, except in E4, are brought down to more reasonable values of 1.5-4 times solar. All the continuum and reflection parameters follow a similar trend as M1, and the presence of the absorption line is also noted. The densities are consistently found to be much higher than the \textsc{relxillCp} value of $10^{15}cm^{-3}$. Despite providing a better fit in general, we still face the problem of the disc densities pegging at the upper limit of $10^{19}cm^{-3}$ throughout E2 to E5. 

\begin{figure*}
\centering
	\includegraphics{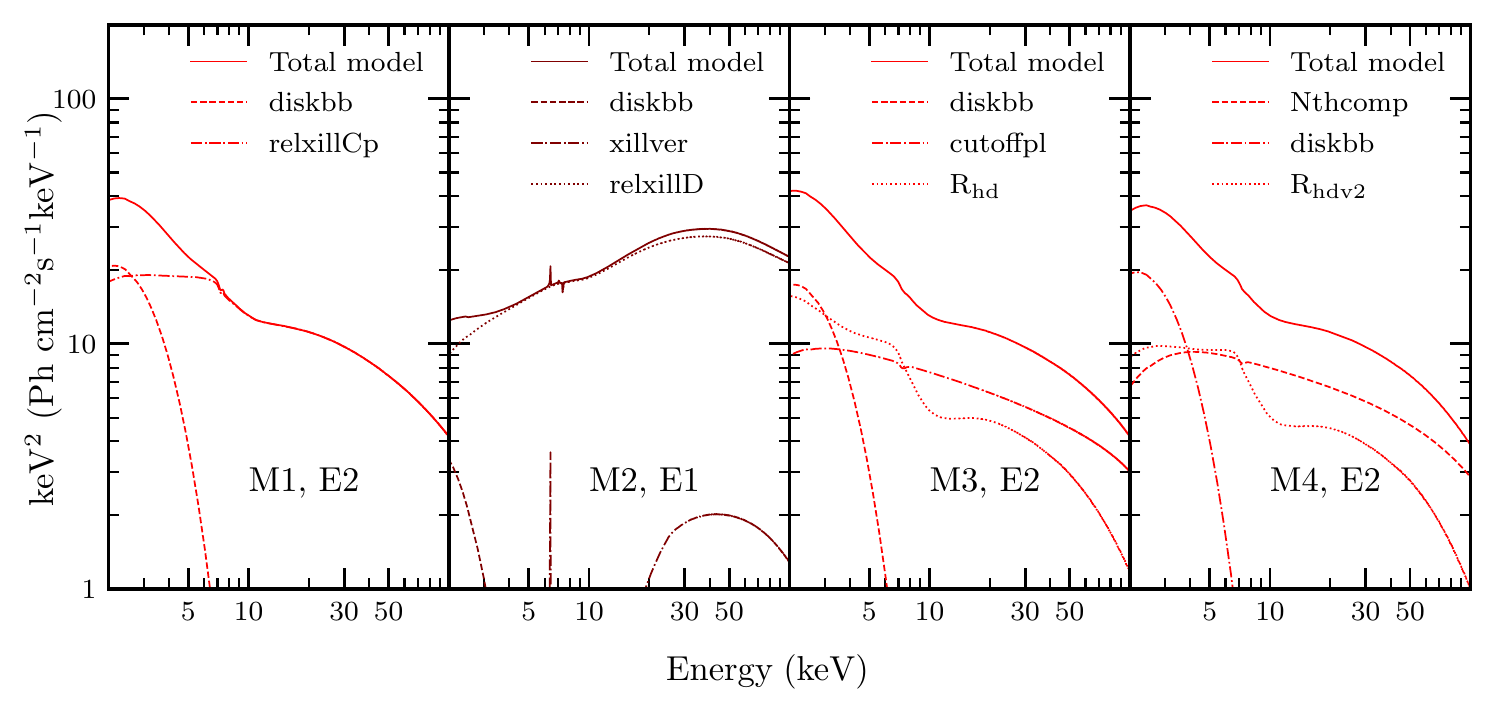}
  	\caption{Best-fit models and model components for all the models (M1,M2,M3,M4), plotted for a representative epoch for each. The model corresponding model and epoch names are displayed in each panel. The bold, dashed, dotted and dash-dotted lines represent the total model, disc, relativistic blurred reflection and the unblurred distant reflection components, respectively. The epochs follow the colour coding from table ~\ref{tab:obs}.
  	 The spectral components are detailed in section ~\ref{sec:w_refl}, ~\ref{sec:relxillD} and ~\ref{sec:reflionxhd}, and the implications discussed in section~\ref{sec:discussion}. }
    \label{fig:eemo}
\end{figure*}

\begin{figure}
\centering
    \includegraphics{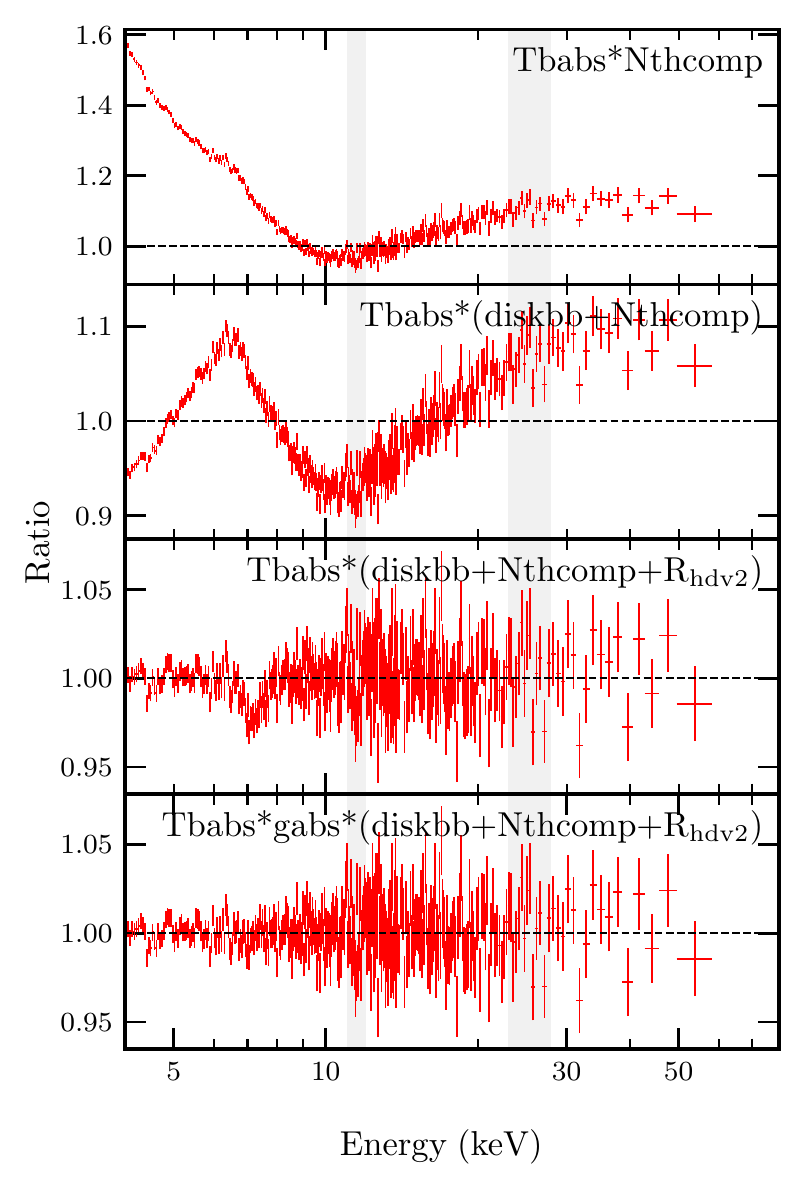}
  	\caption{Residuals, in the form of data/model ratios, for different components of the best-fit model M4 for a representative epochs E2 (colour coded in accordance with table~\ref{tab:obs}). For better visual representation, we display only the FPMA residuals. The panels, from top to bottom, demonstrate the improvement in residuals as further model components are added. $\rm R_{hdv2}$ abbreviates the reflection component, \textsc{relconv}$\otimes$\textsc{reflionx\_hd}.The light grey shaded regions indicate the 11.0-12.0 and 26.0-28.0 keV regions, ignored in this work due to unexplained systematics. See section~\ref{sec:w_refl} for model descriptions.}
    \label{fig:ratio_improvement}
\end{figure}

\subsubsection{reflionx-based high density reflection models} \label{sec:reflionxhd}

While \textsc{relxillD}-based model M2 provides a better fit than M1 and gives an indication of the presence of a high density disc, the disc density is found to peg at its upper limit of $10^{19}cm^{-3}$. To circumvent this issue, we switch from the \textsc{relxill}-based \textsc{relxillD} to \textsc{reflionx}-based \citep{Ross_2005}  high-density reflection model \textsc{relflionx\_hd}, used by \citet{Tomsick_2018} based on the code by \citet{Ross_2007}. \textsc{relflionx\_hd} assumes a cutoff power-law continuum and provides only the reflection component. As opposed to \textsc{relxillD}, \textsc{relflionx\_hd} has the disc density extending upto $10^{22}cm^{-3}$. While similar to \textsc{relxillD}, \textsc{relflionx\_hd} has the cutoff energy fixed at 300 keV, in case of \textsc{relflionx\_hd} the iron abundance is also fixed at the solar value ($A_{\rm Fe}=1$).
We convolve the \textsc{relflionx\_hd} component with smeared relativistic  accretion  disc  line  profiles  using \textsc{relconv} \citep{Dauser_2010}. Thus, for E2-E5, \textsc{relconv*relflionx\_hd} forms the reflection component, while we use \textsc{cutoffpl} as the direct component. Similar to M2, we also use an additional \textsc{xillver} component as required for epoch E3, the cutoff energy of which we also fix at 300 keV.
For the hard state spectra in E1, with low cutoff energy of $<$32 keV, the assumption of $E_{\rm cut}=300$ keV might pose an issue. To take care of this issue, we introduce an exponential cutoff model \textsc{highecut} to the reflection model, and use \textsc{Nthcomp} as the continuum. Following the procedure adapted by \citet{Pintore_2015}, we fix the $E_{{\rm cut}}$ and $E_{{\rm fold}}$ energies in the \textsc{highecut} model to the $kT_{{\rm e}}$ and $2.7\times kT_{{\rm e}}$ of the \textsc{Nthcomp}, respectively. Additionally, we use \textsc{xillverCp} to account for the narrow core of the Fe-K emission. Thus, our model M3 (best-fit parameters detailed in Table ~\ref{tab:M3}) becomes:

Nu02/04:\\
\verb'TBabs'$\times$\verb'gabs'$\times$\verb'(diskbb+Nthcomp+xillverCp+'\\
\verb'(highecut'$\times$\verb'relconv*reflionx_hd))'\\

Nu08:\\
\verb'TBabs'$\times$\verb'gabs'$\times$\verb'(diskbb+cutoffpl+xillver+'\\
\verb'(relconv*reflionx_hd))'\\

Nu06/10/12:\\
\verb'TBabs'$\times$\verb'gabs'$\times$\verb'(diskbb+cutoffpl+(relconv*reflionx_hd))'\\

It can be seen from Table ~\ref{tab:M3} that M3 reduces the $\chi^2$ values as compared to M2, by values between 1 and 46 for E2-E5, depending on the epoch, and by a value of 92 for E1. This improvement over M2 for one less free parameter for E2-E5 and 4 additional free parameters for E1, results in an improvement of the goodness of the fits. The trends in the continuum as similar to before, with the value of $\Gamma$ for E1 being slightly harder than in M1 best-fit and $kT_{\rm e}$ attaining similar values. It is to be noted that the upper limit of the \textsc{cutoffpl} spectral index for \textsc{relflionx\_hd} is 2.3. Thus, in case of E3 where the $\Gamma>2.3$, we fix the $\Gamma$ of the reflection component at 2.3 and allow the $\Gamma$ of the direct \textsc{cutoffpl} component to vary freely. The disc density is found to be much higher than $10^{19}cm^{-3}$, with values ranging in $0.7-27\times10^{20}cm^{-3}$. However, both the disc density and the ionization parameter are found to vary quite significantly between the observations. This could be an artefact of the modeling. 

Although M3 provides a superior fit than M2, \textsc{relflionx\_hd} still has a few limitations. Due to the cutoff energy being fixed at 300 keV, the hard state spectra cannot be consistently modeled with \textsc{relflionx\_hd}. Furthermore, due to the upper limit of spectral index being fixed at 2.3, we are unable to constrain the value properly for E3. Finally, another crucial limitation of \textsc{relflionx\_hd} is that the iron abundance is fixed at the solar value. Super-solar metallicity are quite common in X-ray binaries and are sometimes thought to be caused by radiative levitation \citep{Reynolds_2012}. 

For more flexible modelling, we use a more recent, \textsc{Nthcomp} continuum-based high density reflection model \textsc{reflionx\_hdv2} (John Tomsick, private communication). Here, the electron temperature and the iron
abundance are free parameters, along with the other parameters present in \textsc{reflionx\_hd}. Thus, our model M4 (see Table ~\ref{tab:M4}) becomes:

Nu02/04/08:\\
\verb'TBabs'$\times$\verb'gabs'$\times$\verb'(diskbb+Nthcomp+xillverCp+'\\
\verb'(relconv*reflionx_hdv2))'\\

Nu06/10/12:\\
\verb'TBabs'$\times$\verb'gabs'$\times$\verb'(diskbb+Nthcomp+(relconv*reflionx_hdv2))'\\

From Table ~\ref{tab:M4}, it can be seen that, statistically, M4 provides almost the same quality of fit as M3. The change in $\chi^2$ is minimal, ranging from 1 to 10 for 2 degrees of freedom (1 degree of freedom in case of E1), even though the null hypothesis probability improves. In fact, the AIC and BIC actually increases for E4, while for E3 and E5 they are almost the same. However, as discussed in the beginning of section ~\ref{sec:results}, we have to take into account the consistency of parameter values and their physical interpretation before selecting the optimal model. For M4, the values of $kT_{\rm in}$ and $\Gamma$ are found to be broadly consistent between M4 and M3. Furthermore, the $kT_{\rm e}$ increases monotonically with time from $\sim32-37$ keV to the the maximum value of 400 keV from E1 to E5, while the $R_{\rm in}$ decreases from $\sim 13 R_{\rm g}$ $\sim 2.6 R_{\rm g}$. Apart from E1 (for which the null hypothesis probability is still quite low), all the other inclinations are found to be consistent around 30-38 degrees. As in the case of M3, M4 also shows variations in $\xi$, although the variation is minimal between E2,E3 and E5. Furthermore, the iron abundances are also found to be consistent, hovering around the solar value. Apart form E4, the disc density are constrained to $\sim 10^{20.3-20.7}cm^{-3}$ and are found to be consistent between all the epochs. The normalisations of the \textsc{reflionx\_hdv2} components are found to monotonically decrease between E1 and E5. Finally, the absorption feature around 7.3 keV is found to be prominent and consistent across the epochs. For this consistency and the minimal statistical improvement, we can generally consider M4 as the most appropriate model. It can be observed, however, that for epoch E4, the density assumes a much higher value than the other observations or even compared to best-fit M3 model for E4. M4 also results in a slightly higher BIC and AIC for E4. Thus, even though M3 and M4 are almost equivalent for E4, M3 can as well be considered a more consistent model for E4.

The models M1-M4, along with the individual components, are presented for a few representative epochs in Figure ~\ref{fig:eemo}. The data/model ratios for all the models (M1-M4) are presented for the epochs E3 and E4 in Figure ~\ref{fig:ratio}. The improvement in the fit residuals with the addition of different model components for the best-fit model M4, is presented for a representative epoch E2 in Figure~\ref{fig:ratio_improvement}.

\section{Discussions and Conclusions} \label{sec:discussion}

\begin{figure}
\centering
	\includegraphics{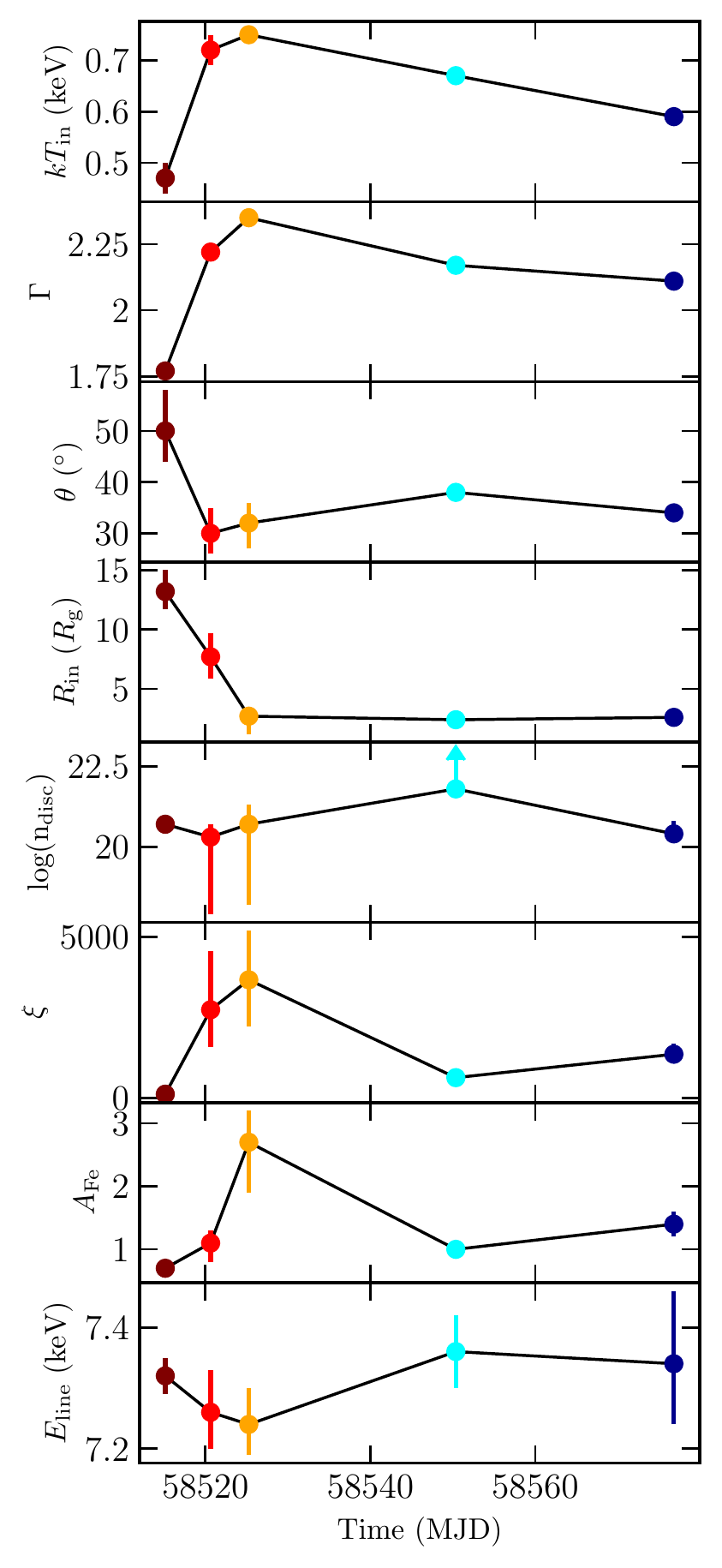}
  	\caption{Time evolution of best-fit parameters for the best-fit model M4. The model is described in section ~\ref{sec:reflionxhd}, and the parameters are detailed in table ~\ref{tab:M4}. The data points are plotted for the different epochs following the colour scheme detailed in table ~\ref{tab:obs}.}
    \label{fig:time_evolution}
\end{figure}

In this work, we present a consistent \textit{NuSTAR} spectral analysis of the recently discovered BHB MAXI~J1348-630, for 5 different epochs spanning hard, hard-soft transition and soft states during its first complete outburst in 2019. We model the \textit{NuSTAR} spectra with the combination of a multi-coloured disc blackbody component, a Comptonisation continuum and \textsc{relxill}-based relativistic reflection components. The resulting model, M1 (see section ~\ref{sec:w_refl} and table ~\ref{tab:M1}) proves quite inadequate to account for all the spectral features. Extremely high iron abundances, hints of high energy excess and low values of null hypothesis probabilities are found. There are a few scenarios that can explain the overabundance and the occasional high energy excesses. 
Presence of a jet, specially in the hard state, could explain some of the extreme parameters \citep{Nowak_2011}. A jet has indeed been observed for MAXI J1348-630 by \citet{Carotenuto_2021}. However, jet models have more free parameters, a comprehensive study of which is beyond our current work. One may try to approximate the presence of jet with the addition of an excess power-law or broken power-law component. However, this approach is found to lead to no further improvement for MAXI~J1348-630. Alternately, presence of a second Comptonization region \citep{Yamada_2013,Basak_2017,Chakraborty_2020} has been proposed as the emission mechanism for MAXI~J1348-630 \citep{FGarcia_2021}. However, addition of a second corona component did not result in any significant improvement in our results. The third scenario, as investigated for other BHBs by \citet{Tomsick_2018,Jiang_2019a}, could be high density reflection. In our current work, we explore this scenario in detail for MAXI~J1348-630.

In section ~\ref{sec:relxillD}, we explore the high density ($n_{\rm e}>10^{15}\rm cm^{-3}$) reflection in MAXI~J1348-630 using the \textsc{relxill}-based \textsc{relxillD} model. The resulting model, M2 (best-fit parameters detailed in table ~\ref{tab:M2}), is found to yield statistically better fits. While M2 mostly resolves the issue with iron overabundance, the density is found to get pegged at the upper limit ($n_{\rm e}=10^{19}\rm cm^{-3}$) of the model. Moreover, due to the electron temperature being fixed at 300 keV, \textsc{relxillD} is found to be inadequate for the hard state epoch E1. To better constrain the density, we then use the \textsc{reflionx}-based high density reflection models (see section ~\ref{sec:reflionxhd}). The resulting models, M3 and M4 (tables ~\ref{tab:M3} and ~\ref{tab:M4} details the best-fit parameter values), generate fits with AIC and BIC less than M1 or M2. The \textsc{cutoffpl} continuum-based \textsc{reflionx\_hd} is limited by the facts that the iron abundance and the electron temperatures are fixed at the solar value and 300 keV, respectively. The latest, \textsc{Nthcomp} continuum-based model \textsc{reflionx\_hdv2} alleviates these limitations. 
Although both M3 and M4 result in similar qualities of fits from a statistical point of view, M4 has more consistent parametrisation across the epochs. M3, however, results in a more consistent density value for the epoch E4, and a more consistent inclination for E1. 
The best-fit parameters for the most appropriate models (tables ~\ref{tab:M3} and ~\ref{tab:M4}), are found to follow the following evolution (figure ~\ref{fig:time_evolution}):
\begin{itemize}
\item The temperature of the innermost part of the accretion disc ($kT_{\rm in}$) first increases to a maximum value of $\sim$0.75 keV around the epoch E3, and then decreases. This trend exactly follows that found by \citet{Tominaga_2020} and \citet{zhang_2020MNRAS.499..851Z}. The temporal evolution of $\Gamma$ is also found to be similar to $kT_{\rm in}$.
\item The truncation radius of the inner accretion disc gets consistently closer to the black hole as the spectral state progresses from hard to soft state.
\item The corona temperature increases from $\sim$32-37 keV to a high value (pegged at the default maximal value of 400 keV for \textsc{Nthcomp}) as the state goes from hard to soft.
\item The inclination, apart from the hard state, is consistent between the observations. The low null hypothesis probability suggests that M4 may not be the most suitable model for hard state epoch E1. For all the other observations, the inclination is found to be $\sim$30-40 degrees.
\item The disc densities are more or less constant at $\sim10^{(20.3-21.4)}\rm cm^{-3}$. This kind of high disc density has previously been found in other BHBs \citep[e.g.,][]{Tomsick_2018,Jiang_2019a}.
\item The abundances are found to be close to the solar value. This is in stark contrast with the \textsc{relxill}-based M1 model. 
\end{itemize}
All these values are further supported by the MCMC analysis (appendix ~\ref{appendix:mcmc}). This consistency of the different parameters and steady improvement in reduced $\chi^2$ for the high-density reflection models, gives further credence to our conclusion. \citet{Garcia_2016}, however, emphasizes that the atomic physics underneath the the reflection models is only known to be accurate up to densities of $\rm \sim10^{19} cm^{-3}$. Hence, more accurate determinations of the rates of atomic transitions and other quantities can have an impact on the high-density reflection models, thereby affecting our results.

As opposed  to  the \textsc{relxillD}, our \textsc{relconv}$\otimes$\textsc{reflionx\_hd} implementation is not fully self-consistent. In case of a real, extended accretion disc, the density should depend on the distance from the central compact object \citep{Svensson_1994}. However, the \textsc{reflionx\_hd} model assumes an optically thick, slab-like atmosphere of a constant density \citep{Ross_2005}. Thus, in our implementation, the X-rays are assumed to be reflected from a single layer at a single photoionization radius ($R_{\rm ion}$, \citet{Mori_2019}). Therefore, following \citet{Mori_2019}, we perform a sanity check for the self-consistency of best-fit M4 model parameters. Assuming an isotropically emitting illuminating source (the corona), the ionisation parameter in \textsc{reflionx\_hd} model can be redefined as $\xi=\frac{L}{nR_{\rm ion}^2}$ (where $L$ is  the illuminating luminosity, $n$ is the hydrogen number density). Now, the electron density($n_{\rm e}$) quoted by \textsc{reflionx\_hd} can be equated to the disc density ($n_{disc}$), as the accretion disc is believed to be primarily composed of hydrogen (this assumption has been followed throughout this work). Furthermore, we can use the 0.1-200 keV unabsorbed luminosity from the Comptonization and the reflection components (thus, disregarding the \textsc{diskbb}) as a proxy for the illuminating luminosity $L$. This is found to be $\sim 1\times10^{37}-3\times10^{38}$ ergs/s, depending on the epoch in consideration. Thus, using the values of $n_{\rm disc}$ and $\xi$ from table ~\ref{tab:M4}, we can calculate $R_{\rm ion}$ as: 
\begin{equation}
\label{R_ion}
R_{\rm ion}=\sqrt{\frac{L^{\rm comp+refl}_{\rm 0.1-200 \ keV}}{n_{\rm disc}\times \xi}}
\end{equation}
On the other hand, we can utilize the inner radius of the accretion disc $R_{\rm in}$ as quoted by the \textsc{relconv} model from table ~\ref{tab:M4}. Therefore, assuming that the corona is located directly above the central black hole, $R_{\rm ion}\gtrsim R_{\rm in}$ can be imposed as a necessary self-consistency condition. Using the values from table ~\ref{tab:M4}, we find that for all our observations, this condition is satisfied within error bars.

The mass of MAXI~J1348-630 has been reported to be $9.1^{+1.6}_{-1.2} \ M_{\odot}$ by \citet{Jana_2020} and $13\pm2 \ M_{\odot}$ by \citet{Tominaga_2020}, further rectified to be $11\pm2 \ M_{\odot}$ by \citet{Lamer_dustRing_2020}. Using equation 3 from \citet{Wang_2018} and the correction factor from \citet{Kubota_1998} for the soft state \textsc{diskbb} normalisation, we find the black hole mass to be $\sim11.6-19.4 \ M_{\odot}$. We note that the calculated mass from the \textit{NuSTAR} \textsc{diskbb} normalisation is likely to be incorrect, due to lack of $<$3 keV spectra in \textit{NuSTAR}. Even though a simultaneous \swift-XRT or \nicer data could be used to estimate the mass more accurately, we leave such a hard+soft X-ray spectral modelling for this source for a future work. For now, we use the black hole mass and distance estimates from \citet{Lamer_dustRing_2020}. Using the Model M4, we find the integrated unabsorbed 0.1-200 keV luminosity of MAXI~J1348-630 to be $1.3\times10^{38},1.2\times10^{38},1.4\times10^{38},6.6\times10^{37}$ and $3.3\times10^{37}$ erg/s, for E1, E2, E3, E4 and E5, respectively. This implies that From E1 to E5, MAXI~J1348-630 was accreting at $6-24\%$ of the Eddington luminosity ($L_{ \rm Edd}$). Using the solution for radiation pressure-dominated disc from \citet{Svensson_1994} \citep[equation 1 in ][]{Garcia_2016}, we can then explain the consistency of density between E1,E2 and E3, and the increase of density for E4. For a radiation-pressure dominated $\alpha$-disc at high accretion rate ($\dot{m}$ much larger than 0.1), $n_{\rm e}\propto \dot{m}^{-2}$. Assuming an accretion efficiency $\epsilon$ to be 20$\%$ \citep{Novikov_1973blho.conf..343N,Agol_2000}, $\dot{m}=\rm L_X/ \epsilon L_{Edd} \approx L_{0.1-200 \ keV}/ \epsilon L_{Edd}$. Using this relation, we find an inverse relation between the X-ray luminosity $L_{\rm X}$ and the disc density. Therefore, for E1, E2 and E3 (where the 0.1-200 keV luminosity is found to be roughly the same), the disc density remains constant. The disc density then increases for E4, where the 0.1-200 keV luminosity decreases by a factor of 2. However, this does not explain the consistency between the densities derived for E5 and that of E1, E2 and E3. 

Finally, another significant highlight of our work is the detection of absorption feature at $\sim$7.3 keV throughout the epochs. Accounting for this feature, either using the multiplicative Gaussian absorption profile \textsc{gabs} or by using a Gaussian emission line profile with negative normalisation, improves the fit by $\Delta \chi^2 \sim 26-166$ (for 3 fewer degrees of freedom), indicating that the absorption feature is significantly detected. The lines are thus found to have equivalent widths (in absolute value) of $\sim$10-20 eV. 
This kind of absorption lines are indication of outflows in the form of ionised accretion disc winds, believed to be ubiquitous in X-ray binaries \citep[e.g., ][]{Ponti_2012} and important to the accretion process \citep{Begelman_1983}. Disc winds in BHBs are generally known to have an equatorial geometry \citep[e.g., ][]{Ponti_2012}. However, recent evidences \citep{Chiang_2012,Xu_2020,Wang_2021} hint towards the existence of ultrafast winds even in non-equatorial directions. Among the prevalent driving mechanisms proposed, the thermally driven winds prefer high inclinations \citep{Tomaru_2020}, while the MHD-driven winds can flow out at all directions \citep{Fukumura_2017}. Therefore, the discovery of such a low-inclination ultrafast outflow as MAXI~J1348-630, hints towards a magnetic origin (Chakraborty et al., in prep).

To summarise, we have characterised the broadband X-ray spectra of the recently-discovered transient black hole X-ray binary MAXI~J1348-630 through systematic investigation of all the \textit{NuSTAR} data during its first complete outburst in 2019. We find consistent continuum and reflection parameters, and the existence of an ultrafast outflow. We also find, using different \textsc{relxill} and \textsc{reflionx}-based reflection models, a significant evidence of reflection off a high density disc.

\begin{table*}
\caption{Parameters of \textit{NuSTAR} fits to \maxi\ spectra with model M1. The model M1 is detailed in section~\ref{sec:w_refl}. Errors represent 90\% confidence intervals.}
\label{tab:M1}
\begin{tabular}{lllcccccccc}
\hline
Spectral Component & Parameter & \multicolumn{5}{c}{Epoch}  \\\cline{3-7}
\\
& & E1 & E2 & E3 & E4 & E5 \\

\hline

\verb'diskbb' & $kT_\mathrm{in}$ (keV) & $0.48_{-0.07}^{+0.04}$ (Nu02) & $0.69_{-0.02}^{+0.02}$ & $0.80_{-0.01}^{+0.01}$ & $0.69_{-0.01}^{+0.01}$ & $0.59_{-0.01}^{+0.01}$ \\
[1ex]
& & $0.47_{-0.08}^{+0.07}$ (Nu04) & ... & ... & ... & ... \\
[1ex]   
    & norm $(\times 10^4)$ & $1.1_{-0.4}^{+1.0}$ (Nu02) & $1.9_{-0.3}^{+0.4}$ & $1.8_{-0.1}^{+0.1}$ & $1.7_{-0.1}^{+0.1}$ & $1.9_{-0.1}^{+0.2}$ \\
    [1ex]
    & & $1.2_{-1.1}^{+0.5}$ (Nu04) & ... & ... & ... & ... \\

&&&&&&&\\

\verb'relxillCp' & $\Gamma$ & $1.70_{-0.01}^{+0.01}$ (Nu02) & $2.22_{-0.01}^{+0.01}$ & $2.36_{-0.01}^{+0.01}$ & $2.18_{-0.01}^{+0.01}$ & $2.09_{-0.02}^{+0.05}$ \\
[1ex]
    & & $1.70_{-0.01}^{+0.01}$ (Nu04) & ... & ... & ... & ... \\
[1ex]
    &$kT_\mathrm{e}$ (keV) & $31_{-3}^{+2}$ (Nu02) & $184_{-34}^{+97}$ & $>201$ & $>282$ & $>178$ \\
[1ex]
    & & $28_{-2}^{+2}$ (Nu04) & ... & ... & ... & ... \\
[1ex]
    & \textit{$\theta$} ($^\circ$) & $42_{-4}^{+4}$ & $37_{-2}^{+2}$ & $42_{-1}^{+1}$ & $35_{-2}^{+2}$ & $31_{-2}^{+2}$ \\
[1ex]
    & $R_\mathrm{in}$ ($R_\mathrm{g}$) & $10.0_{-4.0}^{+2.7}$ & $<2.7$ & $2.4_{-0.2}^{+0.2}$ & $2.9_{-1.3}^{+1.6}$ & $<3.2$ \\
   [1ex]
    & $\log{\xi}$ (log[erg cm/s]) & $3.40_{-0.10}^{+0.08}$  & $4.44_{-0.07}^{+0.06}$ & $4.00_{-0.08}^{+0.05}$ & $4.22_{-0.06}^{+0.08}$ & $4.05_{-0.15}^{+0.05}$ \\
    [1ex]
    & $A_{\rm Fe}$ ($A_{\rm Fe,\odot}$) & $1.4_{-0.4}^{+0.5}$  & $>5.3$ & $4.1_{-0.3}^{+0.4}$ & $>9.6$ & $>6.8$ \\
    [1ex]
    & $\mathcal{R}$  & $0.09$ & $0.18$ & $0.19$ & $0.28$ & $0.52$ \\
    [1ex]
    & norm & $0.18$ (Nu02) & $0.19$ & $0.13$ & $0.03$ & $0.06$ \\
    &  & $0.17$ (Nu04) & ... & ... & ... & ... \\
&&&&&&&\\

\verb'xillverCp'  & norm $(\times 10^{-2})$ & $1.0$ (Nu02) & ... & $0.6$ & ... & ... \\
[1ex]
                &  & $0.7$ (Nu04) & ... & ... & ... & ... \\

&&&&&&&\\

\verb'gabs' & $E_\mathrm{line}$ (keV) & $7.26_{-0.09}^{+0.07}$ & $7.21_{-0.05}^{+0.04}$ & $7.16_{-0.03}^{+0.04}$ & $7.26_{-0.12}^{+0.18}$ & $7.19_{-0.14}^{+0.11}$ \\
[1ex]
   &$\sigma_\mathrm{line}$ (keV) & $0.14_{-0.09}^{+0.08}$ & $0.11_{-0.04}^{+0.05}$ & $<0.2$ & $<0.2$ & $<0.2$ \\
   [1ex]
   & line depth $(\times 10^{-2})$ & $1.24$ & $1.16$ & $1.63$ & $2.83$ & $0.84$ \\

\hline\\

$\chi^2/{\rm d.o.f.}$  & & $2255/2029$ & $1182/1036$ & $884/804$ & $784/722$ & $721/714$ \\
[1ex]
null hypothesis probability  & & $2.6\times10^{-4}$ & $1.0\times10^{-3}$ & $2.6\times10^{-2}$ & $5.3\times10^{-2}$ & $4.2\times10^{-1}$ \\
[1ex]
AIC  & & $2303.3$ & $1209.9$ & $917.6$ & $812.5$ & $749.0$ \\
[1ex]
BIC  & & $2438.3$ & $1279.3$ & $997.6$ & $876.9$ & $813.2$ \\

\hline\\

    & $C_\mathrm{FPMB}$ & $1.03$ & $1.02$ & $1.02$ & $1.01$ & $1.00$ \\
    [1ex]
Unabsorbed flux & 3.0--70.0 keV & $1.13\times10^{-7}$ (Nu02) & $7.49\times10^{-8}$ & $5.59\times10^{-8}$ & $2.0\times10^{-8}$ & $6.6\times10^{-9}$ \\
[1ex]
 & ($\rm erg/cm^2/s$) & $1.13\times10^{-7}$ (Nu04) & ... & ... & ... & ... \\

\hline\\

\end{tabular}
 \begin{flushleft}
 
 \textbf{Note:}
 $T_\mathrm{in}$: Temperature of the inner disc; norm: Normalisation of the corresponding spectral parameter; $\Gamma$: Asymptotic power-law photon index; $T_\mathrm{e}$: Electron temperature of the corona, determining the high energy rollover; $\theta$: Inclination of the inner disc; $R_\mathrm{in}$: Inner disc radius (in units of $R_\mathrm{\rm g}$); $\xi$: Ionisation parameter of the accretion disc, defined as $\xi=L/nR^{2}$, with \textit{L}, \textit{n}, \textit{R} being the ionising luminosity, gas density and the distance to the ionised source, respectively; $A_{\rm Fe}$: Iron abundance, in the units of solar abundance; $\mathcal{R}$: Reflection fraction; $E_{\rm line}$: The central line energy for the Gaussian absorption model; $\sigma_{\rm line}$: line width of the absorption line; $C_\mathrm{FPMB}$: the flux normalisation constant for FPMB (determined by multiplicative \texttt{`constant'} parameter in the spectral models), estimated with respect to the FPMA flux.

\end{flushleft}
\end{table*}

\begin{table*}
\caption{Parameters of \textit{NuSTAR} fits to \maxi\ spectra with model M2. The model M2 is detailed in section~\ref{sec:relxillD}. Errors represent 90\% confidence intervals.}
\label{tab:M2}
\begin{tabular}{lllcccccccc}
\hline
Spectral Component & Parameter & \multicolumn{5}{c}{Epoch}  \\\cline{3-7}
\\
& & E1 & E2 & E3 & E4 & E5 \\

\hline

\verb'diskbb' & $kT_\mathrm{in}$ (keV) & $0.41_{-0.04}^{+0.04}$ (Nu02) & $0.65_{-0.02}^{+0.02}$ & $0.75_{-0.01}^{+0.01}$ & $0.68_{-0.01}^{+0.01}$ & $0.60_{-0.01}^{+0.01}$ \\
[1ex]
& & $0.37_{-0.06}^{+0.07}$ (Nu04) & ... & ... & ... & ... \\
[1ex]   
    & norm $(\times 10^4)$ & $4.2_{-1.9}^{+4.9}$ (Nu02) & $2.4_{-0.4}^{+0.5}$ & $2.6_{-0.2}^{+0.2}$ & $1.9_{-0.1}^{+0.1}$ & $1.8_{-0.1}^{+0.1}$ \\
    [1ex]
    & & $10.6_{-11.9}^{+9.7}$ (Nu04) & ... & ... & ... & ... \\

&&&&&&&\\

\verb'relxillD' & $\Gamma$ & $1.55_{-0.01}^{+0.01}$ (Nu02) & $2.21_{-0.01}^{+0.03}$ & $2.28_{-0.01}^{+0.01}$ & $2.16_{-0.01}^{+0.01}$ & $2.11_{-0.02}^{+0.02}$ \\
[1ex]
    & & $1.55_{-0.01}^{+0.01}$ (Nu04) & ... & ... & ... & ... \\
[1ex]
    & \textit{$\theta$} ($^\circ$) & $23_{-3}^{+3}$ & $36_{-1}^{+2}$ & $24_{-1}^{+2}$ & $26_{-3}^{+2}$ & $31_{-2}^{+1}$ \\
[1ex]
    & $R_\mathrm{in}$ ($R_\mathrm{g}$) & $9.3_{-1.2}^{+1.5}$ & $<4.2$ & $2.7_{-0.3}^{+0.4}$ & $3.1_{-0.5}^{+0.5}$ & $<3.4$ \\
   [1ex]
    & $\log{\xi}$ (log[erg cm/s]) & $3.82_{-0.02}^{+0.01}$  & $4.09_{-0.06}^{+0.13}$ & $4.05_{-0.10}^{+0.09}$ & $4.02_{-0.09}^{+0.14}$ & $3.40_{-0.17}^{+0.09}$ \\
    [1ex]
    & $A_{\rm Fe}$ ($A_{\rm Fe,\odot}$) & $1.6_{-0.1}^{+0.1}$  & $4.0_{-0.9}^{+0.9}$ & $3.4_{-0.4}^{+0.4}$ & $>6.2$ & $3.8_{-0.9}^{+1.5}$ \\
    [1ex]
    & $\log{n_{\rm disc}}$ (log[$\rm cm^{-3}$]) & $17.8_{-0.5}^{+0.2}$  & $>17.9$ & $>18.5$ & $>18.8$ & $>18.8$ \\
    [1ex]
    & $\mathcal{R}$  & $0.40$ & $0.37$ & $0.79$ & $0.28$ & $0.42$ \\
    [1ex]
    & norm & $0.12$ (Nu02) & $0.16$ & $0.06$ & $0.03$ & $0.01$ \\
    &  & $0.12$ (Nu04) & ... & ... & ... & ... \\
&&&&&&&\\

\verb'xillver'  & norm $(\times 10^{-2})$ & $2.0$ (Nu02) & ... & $0.5$ & ... & ... \\
[1ex]
                &  & $1.8$ (Nu04) & ... & ... & ... & ... \\

&&&&&&&\\

\verb'gabs' & $E_\mathrm{line}$ (keV) & $7.40_{-0.41}^{+0.07}$ & $7.27_{-0.05}^{+0.04}$ & $7.12_{-0.05}^{+0.04}$ & $7.29_{-0.04}^{+0.04}$ & $7.16_{-0.05}^{+0.27}$ \\
[1ex]
   &$\sigma_\mathrm{line}$ (keV) & $<0.2$ & $0.17_{-0.06}^{+0.05}$ & $<0.2$ & $0.15_{-0.05}^{+0.07}$ & $<0.2$ \\
   [1ex]
   & line depth $(\times 10^{-2})$ & $0.93$ & $1.53$ & $1.3$ & $1.80$ & $0.63$ \\

\hline\\

$\chi^2/{\rm d.o.f.}$  & & $2348/2030$ & $1164/1036$ & $861/804$ & $761/722$ & $722/714$ \\
[1ex]
null hypothesis probability  & & $1.1\times10^{-5}$ & $3.3\times10^{-3}$ & $7.9\times10^{-2}$ & $1.5\times10^{-1}$ & $4.1\times10^{-1}$ \\
[1ex]
AIC  & & $2356.4$ & $1192.0$ & $895.4$ & $789.2$ & $749.9$ \\
[1ex]
BIC  & & $2485.9$ & $1261.4$ & $975.5$ & $853.6$ & $814.2$ \\

\hline\\

    & $C_\mathrm{FPMB}$ & $1.03$ & $1.02$ & $1.02$ & $1.00$ & $1.00$ \\
    [1ex]
Unabsorbed flux & 3.0--70.0 keV & $1.13\times10^{-7}$ (Nu02) & $7.51\times10^{-8}$ & $5.64\times10^{-8}$ & $2.0\times10^{-8}$ & $6.65\times10^{-9}$ \\
[1ex]
 & ($\rm erg/cm^2/s$) & $1.14\times10^{-7}$ (Nu04) & ... & ... & ... & ... \\
 
\hline\\

\end{tabular}
 \begin{flushleft}
 
 \textbf{Note:}
 $T_\mathrm{in}$: Temperature of the inner disc; norm: Normalisation of the corresponding spectral parameter; $\Gamma$: Asymptotic power-law photon index; $\theta$: Inclination of the inner disc; $R_\mathrm{in}$: Inner disc radius (in units of $R_\mathrm{\rm g}$); $\xi$: Ionisation parameter of the accretion disc, defined as $\xi=L/nR^{2}$, with \textit{L}, \textit{n}, \textit{R} being the ionising luminosity, gas density and the distance to the ionised source, respectively; 
 $A_{\rm Fe}$: Iron abundance, in the units of solar abundance;
 $n_{\rm disc}$: density of the top layer of the accretion disc, assumed to be equal to the corresponding electron number density;
 $\mathcal{R}$: Reflection fraction; 
 $E_{\rm line}$: The central line energy for the Gaussian absorption model; $\sigma_{\rm line}$: line width of the absorption line; $C_\mathrm{FPMB}$: the flux normalisation constant for FPMB (determined by multiplicative \texttt{`constant'} parameter in the spectral models), estimated with respect to the FPMA flux.

\end{flushleft}
\end{table*}

\begin{table*}
\caption{Parameters of \textit{NuSTAR} fits to \maxi\ spectra with model M3. The model M3 is detailed in section~\ref{sec:reflionxhd}. Errors represent 90\% confidence intervals.}
\label{tab:M3}
\begin{tabular}{lllcccccccc}
\hline
Spectral Component & Parameter & \multicolumn{5}{c}{Epoch}  \\\cline{3-7}
\\
& & E1 & E2 & E3 & E4 & E5 \\

\hline

\verb'diskbb' & $kT_\mathrm{in}$ (keV) & $0.47_{-0.20}^{+0.09}$ (Nu02) & $0.69_{-0.02}^{+0.02}$ & $0.76_{-0.01}^{+0.01}$ & $0.67_{-0.02}^{+0.02}$ & $0.60_{-0.02}^{+0.01}$  \\
[1ex]
& & $0.43_{-0.08}^{+0.09}$ (Nu04) & ... & ... & ... & ... \\
[1ex]   
    & norm $(\times 10^4)$ & $0.9_{-0.2}^{+8.1}$ (Nu02) & $1.5_{-0.6}^{+0.9}$ & $2.1_{-0.1}^{+0.1}$ & $2.0_{-0.3}^{+0.4}$ & $1.8_{-0.3}^{+0.4}$ \\
    [1ex]
    & & $1.5_{-0.7}^{+5.1}$ (Nu04) & ... & ... & ... & ... \\

&&&&&&&\\

\verb'Nthcomp' & $\Gamma$ & $1.74_{-0.01}^{+0.01}$ (Nu02) & ... & ... & ... & ... \\
[1ex]
    & & $1.74_{-0.01}^{+0.01}$ (Nu04) & ... & ... & ... & ... \\
[1ex]
&$kT_\mathrm{e}$ (keV) & $31_{-1}^{+1}$ (Nu02) & ... & ... & ... & ... \\
[1ex]
    & & $29_{-1}^{+1}$ (Nu04) & ... & ... & ... & ... \\
&&&&&&&\\
\verb'cutoffpl' & $\Gamma$ & ... & $2.23_{-0.01}^{+0.02}$ & $2.40_{-0.02}^{+0.02}${*} & $2.15_{-0.03}^{+0.02}$ & $2.09_{-0.05}^{+0.05}$ \\
[1ex]
&&&&&&&\\
\verb'relconv'  & \textit{$\theta$} ($^\circ$) & $34_{-1}^{+1}$ & $38_{-3}^{+2}$ & $30_{-12}^{+4}$ & $35_{-4}^{+4}$ & $35_{-4}^{+4}$ \\
[1ex]
    & $R_\mathrm{in}$ ($R_\mathrm{g}$) & $8.6_{-1.4}^{+1.6}$ & $<5.0$ & $3.0_{-0.5}^{+0.4}$ & $2.5_{-0.1}^{+0.1}$ & $<3.2$  \\
   [1ex]
&&&&&&&\\
\verb'reflionx_hd'  & $\xi$ (erg cm/s) & $167_{-51}^{+60}$ & $3652_{-1002}^{+545}$ & $1608_{-395}^{+663}$ & $796_{-202}^{+398}$ & $1301_{-816}^{+842}$  \\
    [1ex]
    & $n_{\rm disc}$ ($\rm \times10^{20}cm^{-3}$) & $10.7_{-7.7}^{+24.7}$ & $0.7_{-0.3}^{+0.7}$ & $9.2_{-2.7}^{+3.2}$ & $27.1_{-14.5}^{+26.1}$ & $3.3_{-1.9}^{+7.5}$  \\
    [1ex]
    & norm & $60$ (Nu02) & $214$ & $66$ & $14$ & $7.4$ \\
    &  & $55$ (Nu04) & ... & ... & ... & ... \\
&&&&&&&\\

\verb'xillver(Cp)'  & norm $(\times 10^{-2})$ & $0.2$ (Nu02) & ...  & $1.8$ & ... & ... \\
[1ex]
                &  & $0.2$ (Nu04) & ... & ... & ... & ... \\

&&&&&&&\\

\verb'gabs' & $E_\mathrm{line}$ (keV) & $7.00_{-0.05}^{+0.04}$ & $7.36_{-0.05}^{+0.05}$ & $7.27_{-0.06}^{+0.07}$ & $7.39_{-0.05}^{+0.05}$ & $7.38_{-0.12}^{+0.15}$ \\
[1ex]
   &$\sigma_\mathrm{line}$ (keV) & $0.27_{-0.03}^{+0.04}$ & $0.26_{-0.06}^{+0.06}$ & $0.13_{-0.08}^{+0.09}$ & $0.25_{-0.08}^{+0.08}$ & $0.26_{-0.11}^{+0.14}$ \\
   [1ex]
   & line depth $(\times 10^{-2})$ & $3.1$ & $2.4$ & $1.1$ & $3.9$ & $3.3$ \\

\hline\\

$\chi^2/{\rm d.o.f.}$  & & $2256/2026$ & $1154/1037$ & $815/805$ & $756/723$ & $720/715$ \\
[1ex]
null hypothesis probability  && $2.3\times10^{-4}$ & $6.4\times10^{-3}$ & $4.0\times10^{-1}$ & $1.9\times10^{-1}$ & $4.4\times10^{-1}$ \\
[1ex]
AIC  & & $2304.5$ & $1179.8$ & $846.9$ & $781.9$ & $746.0$ \\
[1ex]
BIC  & & $2439.6$ & $1244.2$ & $922.2$ & $841.7$ & $805.7$ \\

\hline\\

    & $C_\mathrm{FPMB}$ & $1.03$ & $1.02$ & $1.02$ & $1.01$ & $1.00$ \\
    [1ex]
Unabsorbed flux & 3.0--70.0 keV & $1.13\times10^{-7}$ (Nu02) & $7.5\times10^{-8}$ & $5.6\times10^{-8}$ & $2.0\times10^{-8}$ & $6.7\times10^{-9}$ \\
[1ex]
 & ($\rm erg/cm^2/s$) & $1.13\times10^{-7}$ (Nu04) & ... & ... & ... & ... \\
 
\hline\\

\end{tabular}
 \begin{flushleft}
 
 \textbf{Note:}
 $T_\mathrm{in}$: Temperature of the inner disc; norm: Normalisation of the corresponding spectral parameter; $\Gamma$: Asymptotic power-law photon index; $\theta$: Inclination of the inner disc; $R_\mathrm{in}$: Inner disc radius (in units of $R_\mathrm{\rm g}$); $\xi$: Ionisation parameter of the accretion disc, defined as $\xi=L/nR^{2}$, with \textit{L}, \textit{n}, \textit{R} being the ionising luminosity, gas density and the distance to the ionised source, respectively; 
 $n_{\rm disc}$: density of the top layer of the accretion disc, assumed to be equal to the corresponding electron number density;
 $E_{\rm line}$: The central line energy for the Gaussian absorption model; $\sigma_{\rm line}$: line width of the absorption line; $C_\mathrm{FPMB}$: the flux normalisation constant for FPMB (determined by multiplicative \texttt{`constant'} parameter in the spectral models), estimated with respect to the FPMA flux.
 
\end{flushleft}
\end{table*}

\begin{table*}
\caption{Parameters of \textit{NuSTAR} fits to \maxi\ spectra with model M4. The model M4 is detailed in section~\ref{sec:reflionxhd}. Errors represent 90\% confidence intervals.}
\label{tab:M4}
\begin{tabular}{lllcccccccc}
\hline
Spectral Component & Parameter & \multicolumn{5}{c}{Epoch}  \\\cline{3-7}
\\
& & E1 & E2 & E3 & E4 & E5 \\

\hline

\verb'diskbb' & $kT_\mathrm{in}$ (keV) & $0.47_{-0.03}^{+0.03}$ (Nu02) & $0.72_{-0.03}^{+0.03}$ & $0.75_{-0.01}^{+0.01}$ & $0.67_{-0.01}^{+0.01}$ & $0.59_{-0.01}^{+0.01}$  \\
[1ex]
& & $0.46_{-0.02}^{+0.03}$ (Nu04) & ... & ... & ... & ... \\
[1ex]   
    & norm $(\times 10^4)$ & $0.3_{-0.2}^{+0.5}$ (Nu02) & $1.4_{-0.2}^{+0.2}$ & $2.5_{-0.1}^{+0.2}$ & $2.0_{-0.2}^{+0.2}$ & $2.0_{-0.1}^{+0.2}$ \\
    [1ex]
    & & $0.3_{-0.2}^{+1.5}$ (Nu04) & ... & ... & ... & ... \\

&&&&&&&\\

\verb'Nthcomp' & $\Gamma$ & $1.77_{-0.01}^{+0.01}$ (Nu02) & $2.22_{-0.03}^{+0.03}$ & $2.35_{-0.03}^{+0.03}$ & $2.17_{-0.02}^{+0.01}$ & $2.11_{-0.01}^{+0.02}$ \\
[1ex]
    & & $1.76_{-0.01}^{+0.01}$ (Nu04) & ... & ... & ... & ... \\
[1ex]
    &$kT_\mathrm{e}$ (keV) & $37_{-1}^{+1}$ (Nu02) & $142_{-26}^{+94}$ & $>10$ & $>169$ & $>175$  \\
[1ex]
    & & $32_{-1}^{+2}$ (Nu04) & ... & ... & ... & ... \\
[1ex]
&&&&&&&\\
\verb'relconv'  & \textit{$\theta$} ($^\circ$) & $50_{-6}^{+8}$ & $30_{-4}^{+5}$ & $32_{-5}^{+4}$ & $38_{-1}^{+1}$ & $34_{-1}^{+1}$ \\
[1ex]
    & $R_\mathrm{in}$ ($R_\mathrm{g}$) & $13.2_{-1.5}^{+1.8}$ & $7.7_{-1.8}^{+2.0}$ & $2.7_{-1.5}^{+0.5}$ & $2.4_{-0.4}^{+0.6}$ & $2.6_{-0.3}^{+0.3}$ \\
   [1ex]
&&&&&&&\\
  \verb'reflionx_hdv2'  & $\xi$ (erg cm/s) & $120_{-6}^{+6}$ & $2736_{-1159}^{+1811}$ & $3664_{-1435}^{+1527}$ & $632_{-208}^{+184}$ & $1358_{-280}^{+320}$  \\
    [1ex]
    & $A_{\rm Fe}$ ($A_{\rm Fe,\odot}$) & $0.7_{-0.1}^{+0.1}$ & $1.1_{-0.3}^{+0.2}$ & $2.7_{-0.8}^{+0.5}$ & $1.0_{-0.1}^{+0.1}$ & $1.4_{-0.2}^{+0.2}$   \\
    [1ex]
    & $\log{n_{\rm disc}}$ (log[$\rm cm^{-3}$]) & $20.7_{-0.3}^{+0.3}$ & $20.3_{-2.4}^{+0.4}$ & $20.7_{-2.5}^{+0.6}$ & $>21.8$ & $20.4_{-0.2}^{+0.4}$  \\
    [1ex]
    & norm & $66$ (Nu02) & $27$ & $15$ & $3.6$ & $3.1$ \\
    &  & $64$ (Nu04) & ... & ... & ... & ... \\
&&&&&&&\\

\verb'xillverCp'  & norm $(\times 10^{-2})$ & $1.3$ (Nu02) & ... & $0.8$ & ... & ...\\
[1ex]
                &  & $0.7$ (Nu04) & ... & ... & ... & ... \\

&&&&&&&\\

\verb'gabs' & $E_\mathrm{line}$ (keV) & $7.32_{-0.03}^{+0.03}$ & $7.26_{-0.06}^{+0.07}$ & $7.24_{-0.05}^{+0.06}$ & $7.36_{-0.06}^{+0.06}$ & $7.34_{-0.10}^{+0.12}$  \\
[1ex]
   &$\sigma_\mathrm{line}$ (keV) & $0.26_{-0.03}^{+0.04}$ & $0.22_{-0.08}^{+0.07}$ & $0.13_{-0.01}^{+0.07}$ & $0.23_{-0.06}^{+0.06}$ & $0.22_{-0.09}^{+0.13}$ \\
   [1ex]
   & line depth $(\times 10^{-2})$ & $2.9$ & $1.6$ & $1.2$ & $3.5$ & $1.9$ \\

\hline\\

$\chi^2/{\rm d.o.f.}$  & & $2245/2025$ & $1144/1035$ & $810/803$ & $757/721$ & $719/713$ \\
[1ex]
null hypothesis probability  & & $4.0\times10^{-4}$ & $1.0\times10^{-2}$ & $4.2\times10^{-1}$ & $1.7\times10^{-1}$ & $4.3\times10^{-1}$ \\
[1ex]
AIC  & & $2295.3$ & $1173.6$ & $846.3$ & $786.9$ & $748.6$ \\
[1ex]
BIC  & & $2435.9$ & $1248.0$ & $931.0$ & $855.9$ & $817.4$ \\

\hline\\

    & $C_\mathrm{FPMB}$ & $1.03$ & $1.03$ & $1.03$ & $1.02$ & $1.00$ \\
    [1ex]
Unabsorbed flux & 3.0--70.0 keV & $1.13\times10^{-7}$(Nu02) & $7.5\times10^{-8}$ & $5.6\times10^{-8}$ & $2.0\times10^{-8}$ & $6.7\times10^{-9}$ \\
[1ex]
 & ($\rm erg/cm^2/s$) & $1.13\times10^{-7}$ (Nu04) & ... & ... & ... & ... \\
 
\hline\\

\end{tabular}
 \begin{flushleft}
 
 \textbf{Note:}
 $T_\mathrm{in}$: Temperature of the inner disc; norm: Normalisation of the corresponding spectral parameter; $\Gamma$: Asymptotic power-law photon index; $\theta$: Inclination of the inner disc; $R_\mathrm{in}$: Inner disc radius (in units of $R_\mathrm{\rm g}$); $\xi$: Ionisation parameter of the accretion disc, defined as $\xi=L/nR^{2}$, with \textit{L}, \textit{n}, \textit{R} being the ionising luminosity, gas density and the distance to the ionised source, respectively; 
 $A_{\rm Fe}$: Iron abundance, in the units of solar abundance;
 $n_{\rm disc}$: density of the top layer of the accretion disc, assumed to be equal to the corresponding electron number density;
 $E_{\rm line}$: The central line energy for the Gaussian absorption model; $\sigma_{\rm line}$: line width of the absorption line; $C_\mathrm{FPMB}$: the flux normalisation constant for FPMB (determined by multiplicative \texttt{`constant'} parameter in the spectral models), estimated with respect to the FPMA flux.

\end{flushleft}
\end{table*}

\section*{Acknowledgements}

We thank the referee for constructive comments which improved the paper. This research has made use of the $MAXI$ data provided by RIKEN, JAXA and the $MAXI$ team. This research has also made use of the \textit{NuSTAR} Data Analysis Software (NuSTARDAS), jointly developed by the ASI Science Data Center (ASDC, Italy) and the California Institute of Technology (USA). 
The authors also thank Samuzal Barua, Vikas Chand, and Mihoko Yukita for valuable insights and comments regarding data reduction and analysis. We thank Michael Parker for constructing the \textsc{reflionx\_hd} and \textsc{reflionx\_hdv2} table models. JAT acknowledges partial support from NASA ADAP grant 80NSSC19K0586.

\section*{Data Availability}

The observational data used in this paper are publicly available at NASA's High Energy Astrophysics Science Archive Research Center (HEASARC; \url{https://heasarc.gsfc.nasa.gov/}). All the spectral figures are generated using the python implementation (pyxspec) of XSPEC. The MCMC results are plotted using the cornner.py module \citep{corner}. Any additional information will be available upon reasonable request.

\bibliographystyle{mnras}
\bibliography{maxi_j1348}


\appendix

\section{MCMC results}\label{appendix:mcmc}

\begin{figure}
\centering
	\includegraphics[angle=0.0,width=1.0\linewidth]{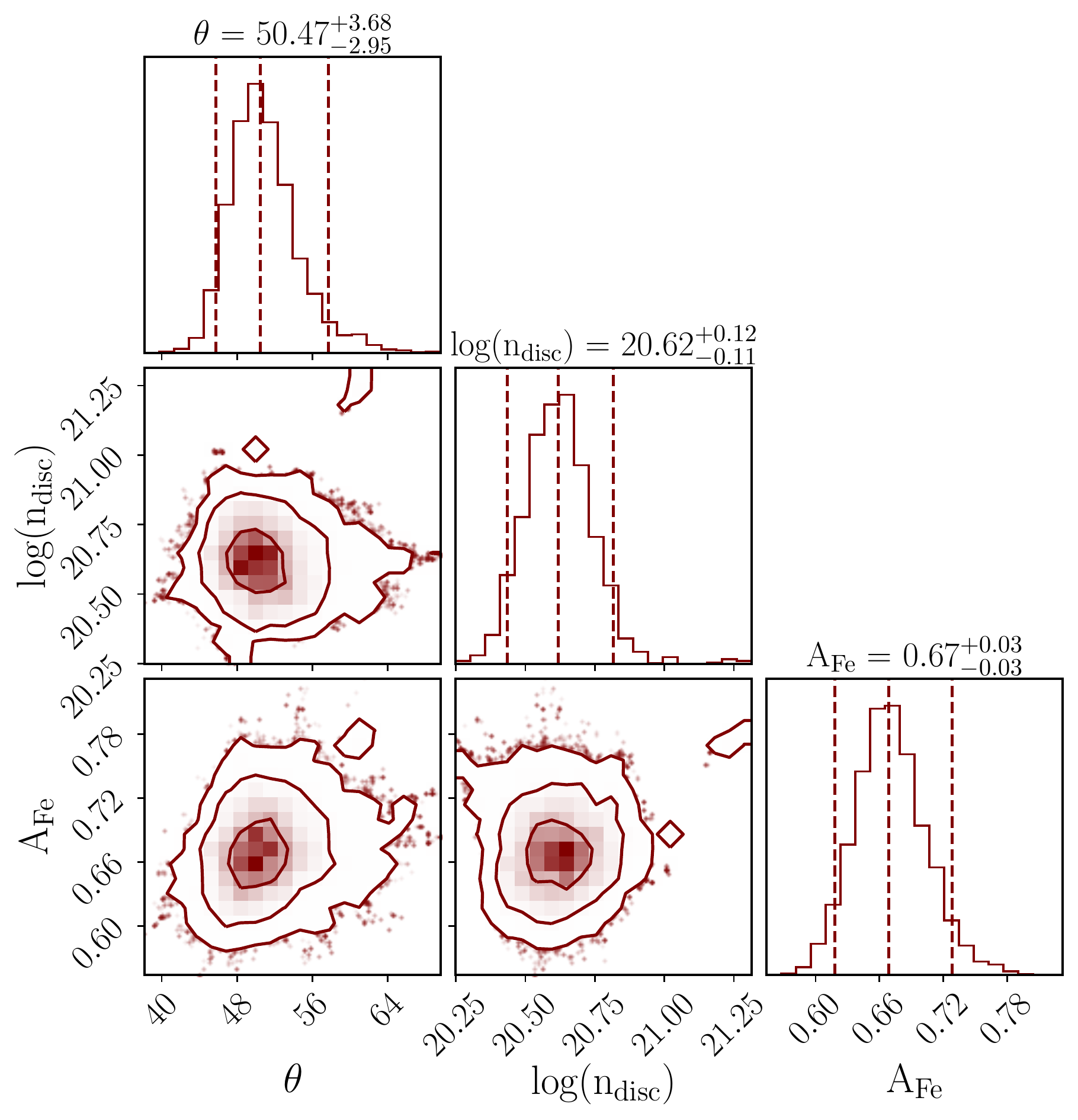}
  	\caption{MCMC results for the epoch E1 and best-fit model M4.}
    \label{fig:mcmc1}
\end{figure}

\begin{figure}
\centering
	\includegraphics[angle=0.0,width=1.0\linewidth]{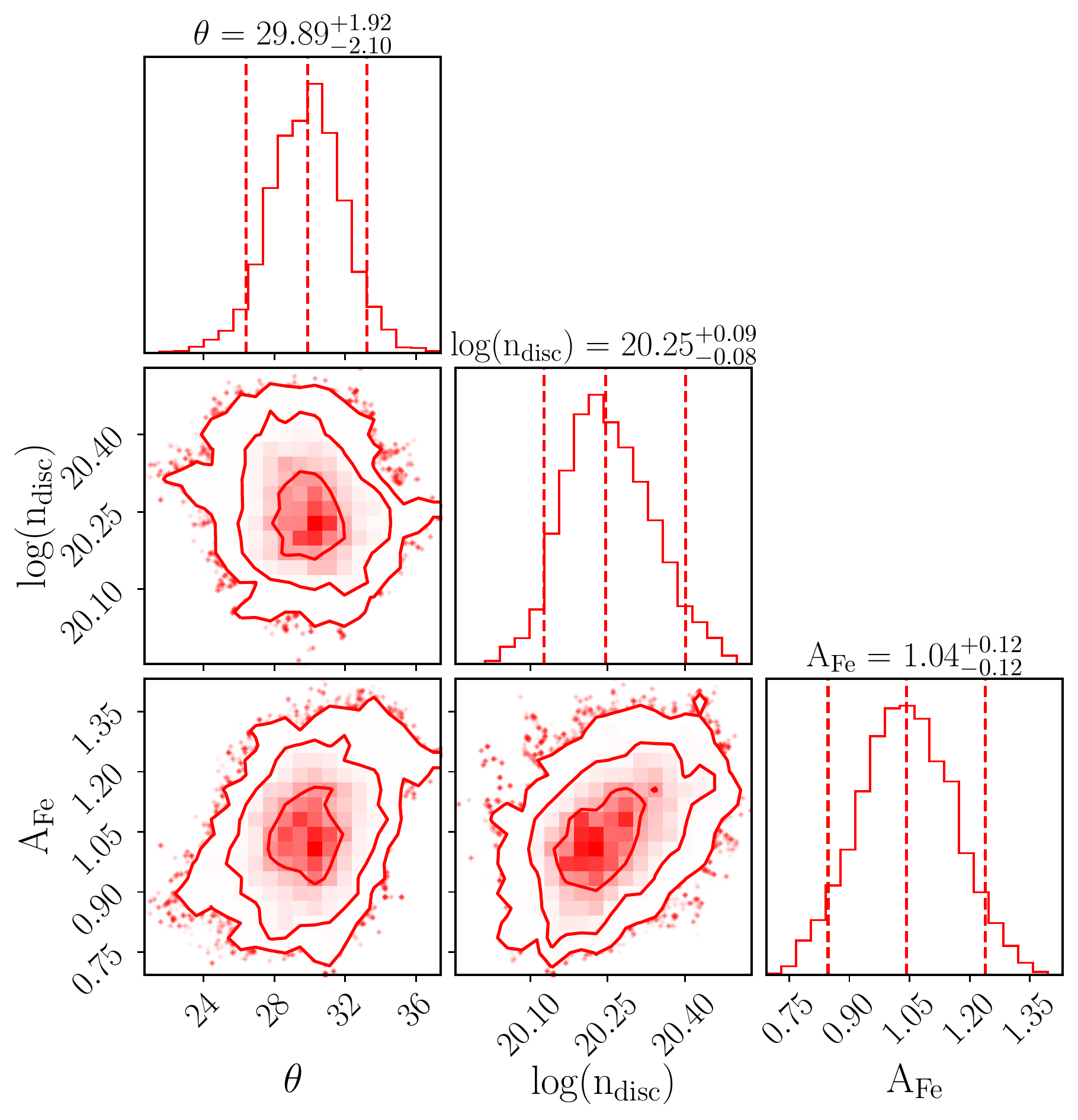}
  	\caption{MCMC results for the epoch E2 and best-fit model M4.}
    \label{fig:mcmc2}
\end{figure}

\begin{figure}
\centering
	\includegraphics[angle=0.0,width=1.0\linewidth]{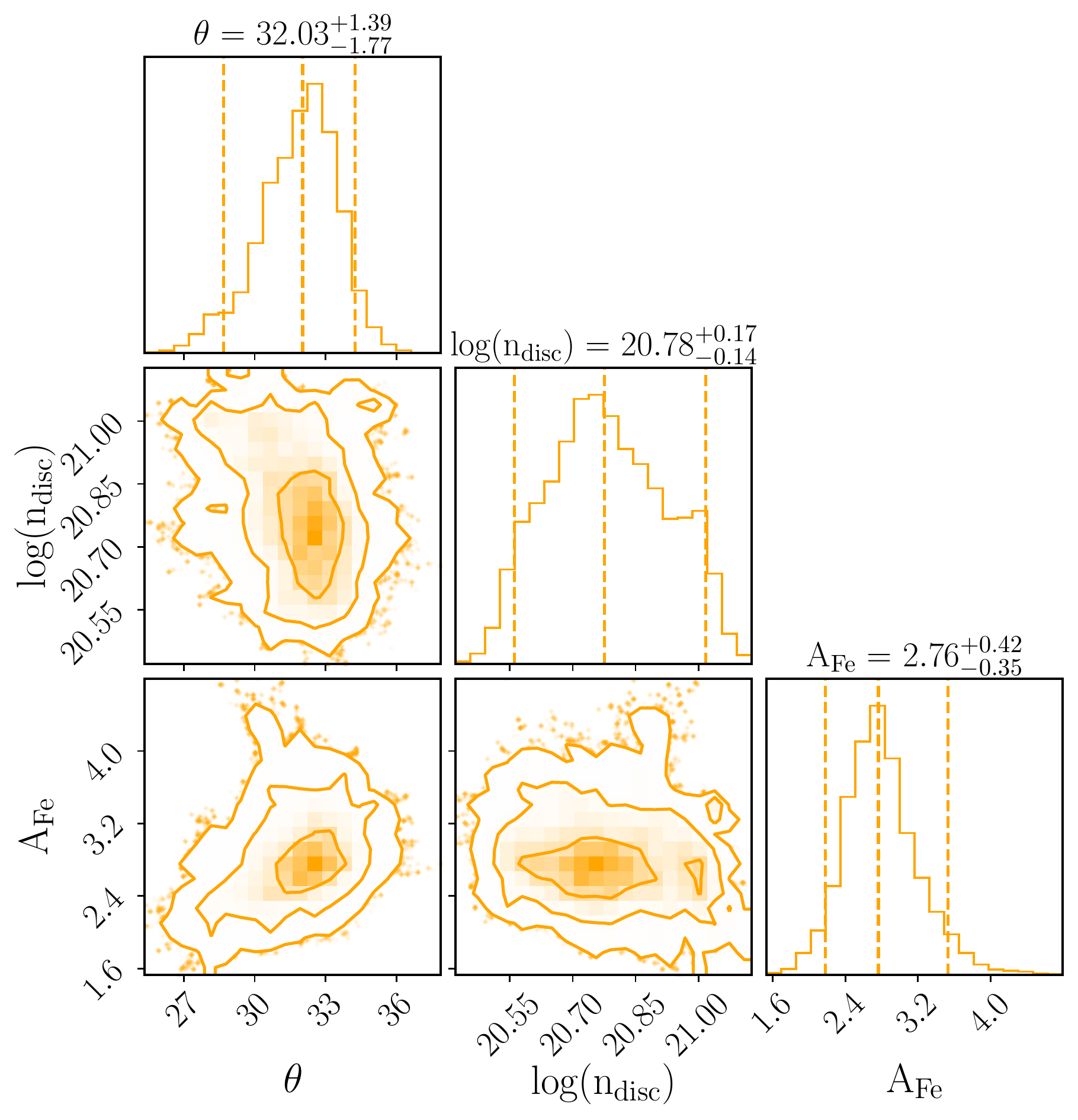}
  	\caption{MCMC results for the epoch E3 and best-fit model M4.}
    \label{fig:mcmc3}
\end{figure}

\begin{figure}
\centering
	\includegraphics[angle=0.0,width=1.0\linewidth]{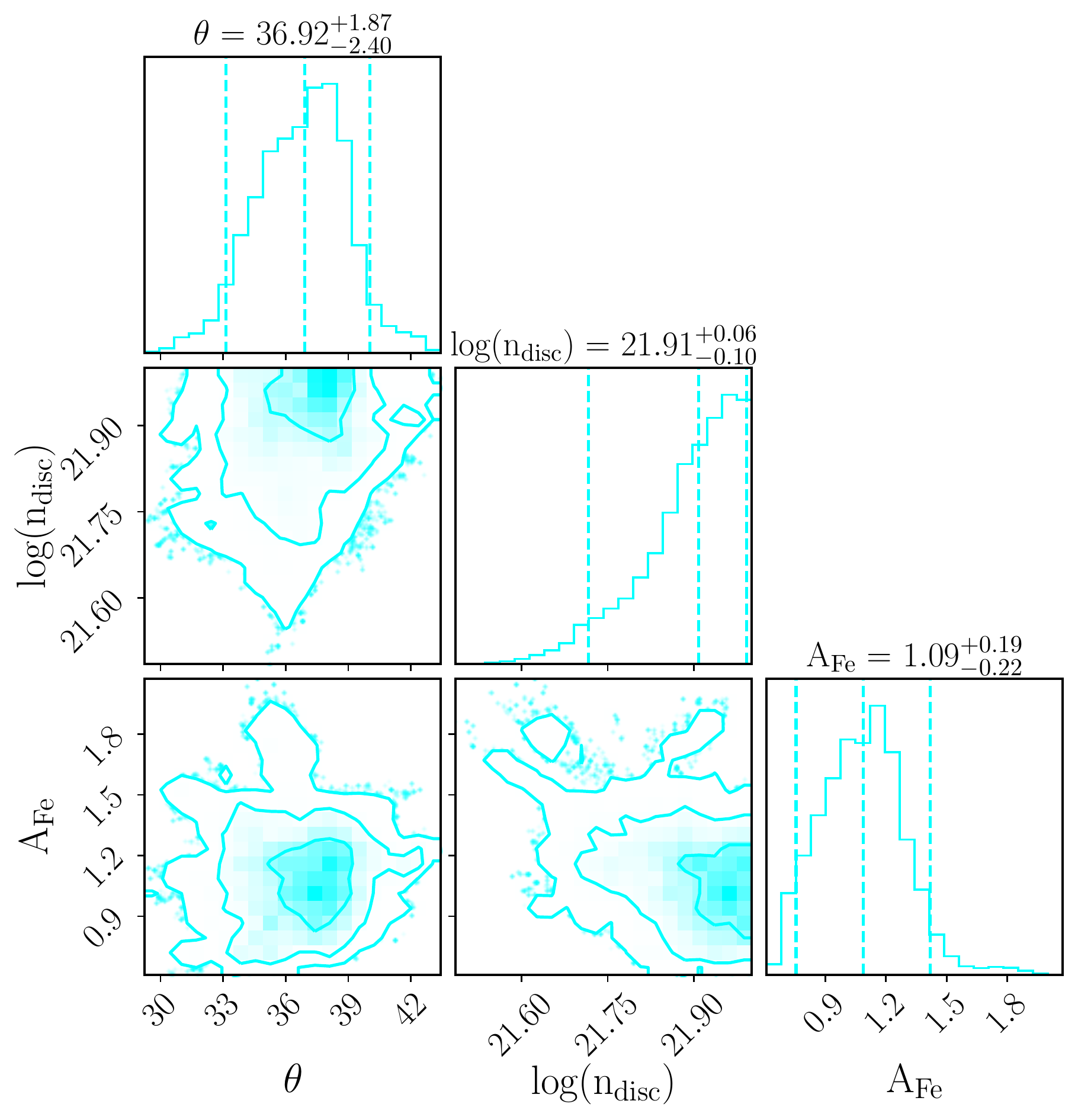}
  	\caption{MCMC results for the epoch E4 and best-fit model M4.}
    \label{fig:mcmc4}
\end{figure}

\begin{figure}
\centering
	\includegraphics[angle=0.0,width=1.0\linewidth]{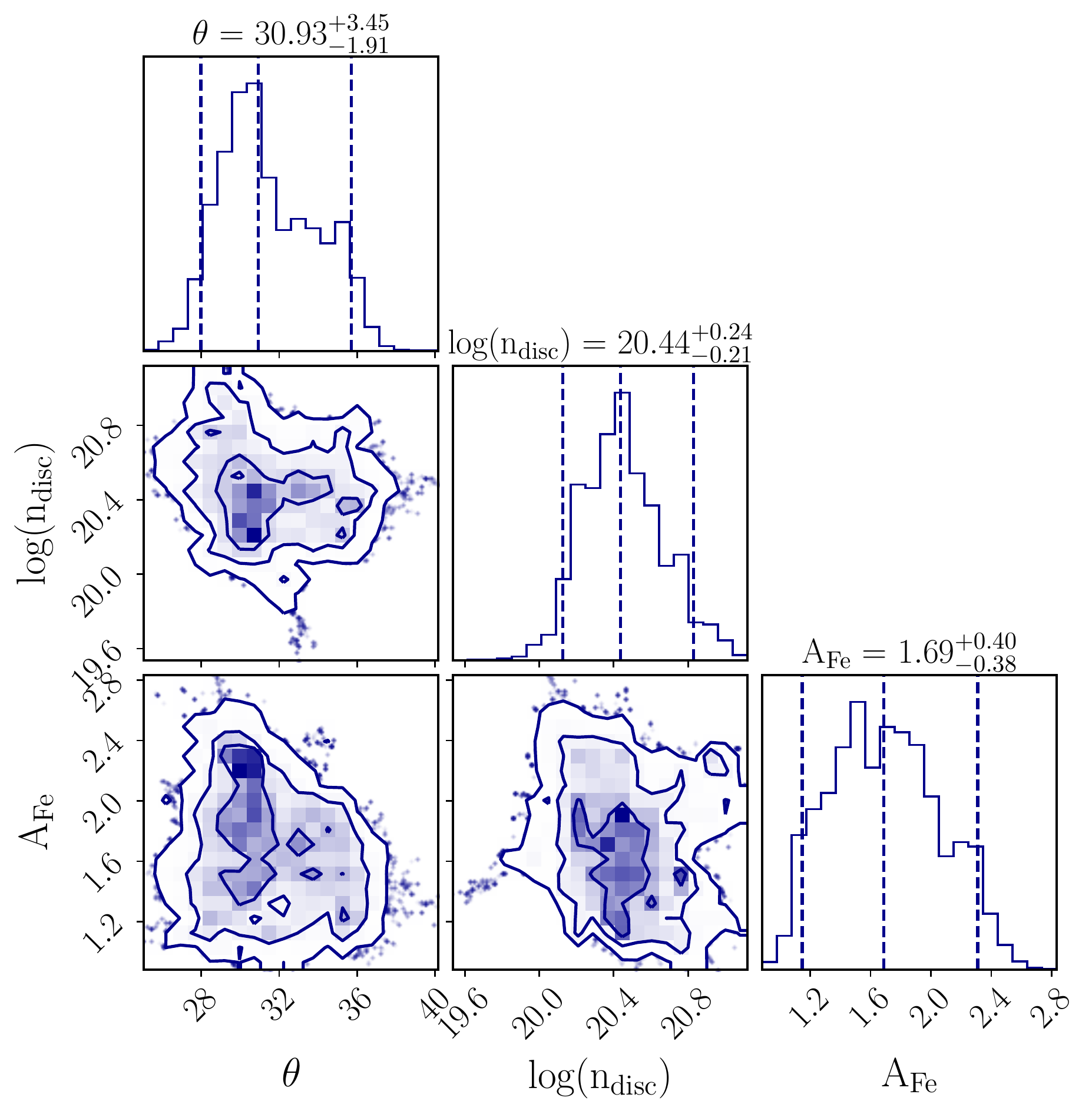}
  	\caption{MCMC results for the epoch E5 and best-fit model M4.}
    \label{fig:mcmc5}
\end{figure}

To estimate better errorbars for some of the most important quantities for out model M4, we perform a Markov chain Monte Carlo (MCMC) analysis by using the XSPEC implementation of the EMCEE code (\textsc{xspec\_emcee}\footnote{\url{https://github.com/jeremysanders/xspec_emcee}}), written by Jeremy Sanders. We use the Goodman–Weare algorithm, with 50 walkers, chain lengths of 5000, and burn-in lengths of 500. The results of the MCMC are plotted in figures ~\ref{fig:mcmc1},~\ref{fig:mcmc2},~\ref{fig:mcmc3},~\ref{fig:mcmc4} and ~\ref{fig:mcmc5}. The contours denote the 1,2 and 3-$\sigma$ confidence contours for the 2D posterior distributions, while the value and errors quoted are the mean and 68\% confidence levels for the 1D posterior distributions. The values are consistent with our spectral fits and gives further credence to our results.

\bsp	
\label{lastpage}
\end{document}